\documentclass[aps,prd,onecolumn,showpacs,nofootinbib,superscriptaddress]{revtex4}  
\usepackage{graphicx}
\usepackage{epstopdf}
\usepackage{amsmath}
\usepackage{amsfonts}
\usepackage{amssymb}
\usepackage{appendix}
\usepackage{comment}
\usepackage{bbold}
\usepackage{color}
\usepackage{slashed}
\usepackage{subfigure}
\usepackage{setspace}
\usepackage{footnote}
\usepackage{multirow}

\begin{document}

\singlespacing

{\hfill NUHEP-TH/15-02}

\title{A Sterile Neutrino at DUNE}

\author{Jeffrey M. Berryman}
\affiliation{Northwestern University, Department of Physics \& Astronomy, 2145 Sheridan Road, Evanston, IL 60208, USA}
\author{Andr\'{e} de Gouv\^{e}a} 
\affiliation{Northwestern University, Department of Physics \& Astronomy, 2145 Sheridan Road, Evanston, IL 60208, USA}
\author{Kevin J. Kelly}
\affiliation{Northwestern University, Department of Physics \& Astronomy, 2145 Sheridan Road, Evanston, IL 60208, USA}
\author{Andrew Kobach}
\affiliation{Northwestern University, Department of Physics \& Astronomy, 2145 Sheridan Road, Evanston, IL 60208, USA}
\affiliation{Department of Physics, University of California, San Diego, La Jolla, CA 92093, USA}

\begin{abstract}
We investigate the potential for the Deep Underground Neutrino Experiment (DUNE) to probe the existence and effects of a fourth neutrino mass-eigenstate. We study the mixing of the fourth mass-eigenstate with the three active neutrinos of the Standard Model, including the effects of new sources of $CP$-invariance violation, for a wide range of new mass-squared differences, from lower than $10^{-5}$~eV$^2$ to higher than 1~eV$^2$. DUNE is sensitive to previously unexplored regions of the mixing angle -- mass-squared difference parameter space. If there is a fourth neutrino, in some regions of the parameter space, DUNE is able to measure the new oscillation parameters (some very precisely) and clearly identify two independent sources of $CP$-invariance violation. Finally, we use the hypothesis that there are four neutrino mass-eigenstates in order to ascertain how well DUNE can test the limits of the three-massive-neutrinos paradigm. In this way, we briefly explore whether light sterile neutrinos can serve as proxies for other, in principle unknown, phenomena that might manifest themselves in long-baseline neutrino oscillation experiments.
\end{abstract}


\maketitle

\setcounter{equation}{0}
\section{Introduction}
\label{introduction}

It is now established, beyond reasonable doubt, that neutrinos have mass and that leptons mix. In order to further explore the neutrino sector and exploit the oscillation phenomenon, ambitious next-generation, long-baseline neutrino oscillation experiments are under serious consideration, including the Deep Underground Neutrino Experiment (DUNE) proposal in the United States \cite{Adams:2013qkq} and the HyperKamiokande detector (and accompanying neutrino source in J-PARC) in Japan \cite{Kearns:2013lea}. The goals of these projects include the search for leptonic $CP$-invariance violation and testing the limits of the so-called three-massive-neutrinos paradigm (for more see, for example, Ref.~\cite{deGouvea:2013onf}).  

The hypothesis that there are three neutrinos, at least two of them massive, and that these interact as prescribed by the Standard Model of electroweak interactions, accommodates almost all neutrino data. There is, however, plenty of room for new phenomena. The unitarity of the leptonic mixing matrix, for example, has not been thoroughly explored\footnote{Arguably, the data only allow unitarity checks for the first row, $\sum_{i=1,2,3}|U_{ei}|^2\stackrel{?}{=}1$, and third column, $\sum_{\alpha=e,\mu,\tau}|U_{\alpha 3}|^2\stackrel{?}{=}1$.} (see, for example, Ref.~\cite{Qian:2013ora}). In spite of tremendous experimental progress, little nontrivial information regarding the appropriateness of the three-massive-neutrinos paradigm has been collected over the past seventeen years. 

Next-generation long-baseline experiments can probe several different new phenomena, including new, weaker-than-weak, interactions involving neutrinos and charged-fermions \cite{Ohlsson:2012kf} that lead to anomalous matter effects, new long-range forces \cite{GonzalezGarcia:2007ib}, and the existence of very light new states \cite{Abazajian:2012ys}.  Here, we concentrate on the simple hypothesis that there is a fourth neutrino mass-eigenstate of unknown, but very small, mass -- $m_4$ less than a few eV -- and assume that there is a nonzero probability that the new neutrino state can be measured as one of the active neutrino states, i.e., we assume the leptonic mixing matrix $U$ to be $4\times 4$ and that $U_{\alpha 4}$, $\alpha=e,\mu,\tau$ are nonzero. Electroweak precision data require the fourth neutrino flavor-eigenstate not to interact with the $W$ and $Z$ bosons with Standard Model strength, so we refer to it as a sterile neutrino. 

We concentrate on this hypothesis for a few different reasons. It is simple and easy to parameterize, and very familiar \cite{Abazajian:2012ys}. Indeed, certain aspects of the effects of sterile neutrinos on different long-baseline experiments have been studied in the recent past (see, for example, Refs.~\cite{Donini:2007yf,Dighe:2007uf,deGouvea:2008qk,Meloni:2010zr,Bhattacharya:2011ee,Hollander:2014iha,Klop:2014ima}). Sterile neutrinos are also a very natural and benign extension of the Standard Model and could be a side effect of the mechanism responsible for the nonzero neutrino masses (see, for example, Refs.~\cite{deGouvea:2005er,Asaka:2005an}). Furthermore, there are the so-called short-baseline anomalies \cite{Aguilar:2001ty,AguilarArevalo:2008rc,Mention:2011rk,Frekers:2011zz,Aguilar-Arevalo:2013pmq}. These might be pointing to more new physics in the leptonic sector, but a convincing, robust explanation remains elusive. Sterile neutrino interpretations to the short-baseline anomalies are, arguably, the simplest explanations of these data. Here, we remain agnostic regarding the new-physics origin of the short-baseline anomalies but, on occasion, will highlight the region of mass and mixing space that is preferred by them. Finally, we would like to explore whether light sterile neutrinos can serve as proxies for other, in principle unknown, phenomena that might manifest themselves in long-baseline neutrino oscillation experiments. This aspect of our analysis will become more clear later. 

This paper is organized as follows. In Section~\ref{parameterization}, we review four-flavor neutrino oscillations. Since we are interested in a large range of new mass-squared differences, we pay special attention to the neutrino oscillation probabilities in the limits when the new oscillation frequency is much larger or much smaller than the known oscillation frequencies. In Section~\ref{DUNE}, we discuss in detail the capabilities of DUNE to (a) rule out the sterile neutrino hypothesis assuming the data are consistent with the three-massive-neutrinos paradigm; (b) determine the new mixing parameters assuming there is one sterile neutrino. Here, we assume new mass-squared differences that range from $10^{-5}$~eV$^2$ to $1$~eV$^2$; and (c)  diagnose that there is physics beyond the three-massive-neutrinos paradigm assuming there is a fourth mass-eigenstate. In Section  \ref{conclusions}, we offer some concluding remarks.

%%%%%%%%%%%%%%%%%%%%%%%%%%%%%
\setcounter{equation}{0}
\section{Oscillations With Four Neutrinos}
\label{parameterization}
%%%%%%%%%%%%%%%%%%%%%%%%%%%%%

We consider a fourth neutrino $\nu_s$ that does not participate in the weak interactions but that can mix with the other three neutrinos of the Standard Model. The misalignment between the mass-eigenstates $\nu_i$ ($i=1$, $2$, $3$, $4$) and flavor-eigenstates $\nu_\alpha$ ($\alpha=e$, $\mu$, $\tau$, $s$) can be described by a general $4\times4$ unitary matrix parameterized by six angles $\phi_{ij}$ ($i$, $j = 1$, $2$, $3$, $4$; $i< j$) and three phases $\eta_1$, $\eta_2$, $\eta_3$. We choose the matrix elements $U_{\alpha i}$ to be:\footnote{We ignore potential Majorana phases because they do not affect oscillations in any realistically observable way.} 
\begin{align}
U_{e2} \text{ } = &  \text{ } s_{12} c_{13} c_{14}, \\
U_{e3} \text{ } = & \text{ } e^{-i \eta_1} s_{13} c_{14},  \\
U_{e4} \text{ } = & \text{ } e^{-i \eta_2} s_{14},  \\
U_{\mu2} \text{ } = & \text{ } c_{24} \left(c_{12} c_{23} - e^{i \eta_1} s_{12} s_{13} s_{23} \right) - e^{i(\eta_2 - \eta_3)} s_{12} s_{14} s_{24} c_{13} ,  \\
U_{\mu3} \text{ } = & \text{ } s_{23} c_{13} c_{24} - e^{i(\eta_2 - \eta_3 -\eta_1)} s_{13} s_{14} s_{24}, \\
U_{\mu4} \text{ } = &\text{ } e^{-i \eta_3} s_{24} c_{14},  \\
U_{\tau2} \text{ } = & \text{ } c_{34} \left(-c_{12} s_{23} - e^{i \eta_1} s_{12} s_{13} c_{23} \right) - e^{i\eta_2} c_{13} c_{24} s_{12} s_{14} s_{34} \nonumber \\
&  \text{ } - e^{i \eta_3} \left(c_{12} c_{23} - e^{i \eta_1} s_{12} s_{13} s_{23} \right) s_{24} s_{34}, \\
U_{\tau3} \text{ } = & \text{ } c_{13} c_{23} c_{34} - e^{i(\eta_2 -\eta_1)} s_{13} s_{14} s_{34} c_{24} - e^{i \eta_3} s_{23} s_{24} s_{34} c_{13}, \\
U_{\tau4} \text{ } = &\text{ } s_{34} c_{14} c_{24},
\end{align}
where $s_{ij} \equiv \sin \phi_{ij}$, $c_{ij} \equiv \cos \phi_{ij}$. The matrix elements not listed here can be determined through the unitarity conditions of $U$. 

When the new mixing angles $\phi_{14}$, $\phi_{24}$, and $\phi_{34}$ vanish, one encounters oscillations among only three neutrinos, and we can map the remaining parameters $\{\phi_{12}$, $\phi_{13}$, $\phi_{23}$, $\eta_1\} \to \{\theta_{12}$, $\theta_{13}$, $\theta_{23}$, $\delta_{CP}\}$, the well-known mixing parameters that define the standard $3\times3$ leptonic mixing matrix in the three-massive-neutrinos paradigm (using the Particle Data Group convention~\cite{Agashe:2014kda}). In the limit where $\phi_{14}$, $\phi_{24}$, and $\phi_{34}$ are small, the angles $\phi_{12}$, $\phi_{13}$ and $\phi_{23}$ play roles very similar to those of $\theta_{12}$, $\theta_{13}$ and $\theta_{23}$, respectively. We discuss this in more detail in Section~\ref{DUNE}. The best-fit values from Ref.~\cite{Agashe:2014kda} for a three-flavor fit to existing data are $\sin^2 \theta_{12} = 0.308 \pm 0.017$, $\sin^2 \theta_{13} = 0.0234^{+0.0020}_{-0.0019}$, and $\sin^2 \theta_{23} = 0.437^{+0.033}_{-0.023}$; the $CP$-odd phase $\delta_{CP}$ is virtually unconstrained.

The amplitude for $\nu_\alpha$ to be detected as $\nu_\beta$ after propagating a distance $L$ in vacuum is
\begin{equation}
\label{oscamp}
\mathcal{A}_{\alpha\beta} = \delta_{\alpha\beta} + U_{\alpha 2}U_{\beta 2}^* \left( e^{-i \Delta_{12}}-1 \right) + U_{\alpha 3}U_{\beta 3}^* \left( e^{-i \Delta_{13}}-1 \right) + U_{\alpha 4}U_{\beta 4}^* \left( e^{-i \Delta_{14}} -1 \right),
\end{equation}
where $\Delta_{ij} \equiv 2.54 \left( \Delta m^2_{ij}/1 \text{ eV}^2\right) \left( L/1 \text{ km} \right) \left( 1 \text{ GeV}/E_\nu \right)$,  $E_\nu$ is the neutrino energy, $\Delta m^2_{ij} \equiv m^2_j - m^2_i$, and $m_i$ is the mass of $\nu_i$. The corresponding probability is $P_{\alpha\beta} = \vert \mathcal{A}_{\alpha\beta} \vert^2$. Eq.~(\ref{oscamp}) assumes that the four mass-eigenstates remain coherent over the neutrino's evolution.\footnote{If $\nu_4$ decoheres from the other three neutrinos, then the expression for $P_{\alpha\beta}$ is modified by neglecting the interference of the oscillations related to $\Delta_{14}$ with those of $\Delta_{12}$ and $\Delta_{13}$, cf.~Eq.~(\ref{prob_ave}). Decoherence occurs if, for example, $\nu_4$ is produced incoherently, or, during propagation, the $\nu_4$ wavepacket becomes well-separated from the wavepacket containing $\nu_1$, $\nu_2$ and $\nu_3$. 
}
The amplitude $\mathcal{A}_{\overline{\alpha\beta}}$ for $\overline{\nu}_{\alpha}$ to be detected as $\overline{\nu}_{\beta}$ is equal to $\mathcal{A}_{\alpha\beta}$ in Eq.~(\ref{oscamp}) with the exchange $U_{\alpha i}U^*_{\beta i} \leftrightarrow U^*_{\alpha i}U_{\beta i}$, for all $i=2,3,4$. Unless otherwise noted, we will assume that the values of the mass-squared splittings $\Delta m^2_{12}$ and $\Delta m^2_{13}$ are very close to the ones that fit the neutrino data assuming there are only three neutrino species (explicitly, $\Delta m_{12}^2 = 7.54\times 10^{-5}$ eV$^2$, $\Delta m_{13}^2 = 2.43\times 10^{-3}$ eV$^2$, assuming the neutrino mass hierarchy is normal~\cite{Agashe:2014kda}), as we will discuss in Sec.~\ref{DUNE}. The value of $m_4$ is mostly unconstrained, so $\Delta m^2_{14}$ can be larger or smaller than $\Delta m^2_{12}$ and $\Delta m^2_{13}$. We do, however, restrict our analyses to positive $\Delta m^2_{14}$, i.e., $m_4 > m_1$. In summary, including the fact that we will always assume the normal neutrino mass hierarchy for the mostly active states, our masses are ordered as follows: $m_1<m_2<m_3$, and $m_4>m_1$. As we vary $\Delta m^2_{14}$ from very small to very large, we allow for all different mass orderings: $m_4\le m_{2}<m_3$; $m_2< m_4 < m_{3}$; and $m_2<m_{3}\le m_4$.

The amplitude simplifies considerably when $\Delta_{14} \ll 1$. In this limit, the last term in Eq.~(\ref{oscamp}) is small compared to the others, so  
\begin{equation}
\label{small14}
P_{\alpha\beta} \simeq \bigg\vert  \delta_{\alpha\beta} + U_{\alpha 2}U_{\beta 2}^* \left( e^{-i \Delta_{12}}-1 \right) + U_{\alpha 3}U_{\beta 3}^* \left( e^{-i \Delta_{13}}-1 \right) \bigg\vert^2.
\end{equation}
Because the experimental normalization uncertainties we will consider are $\mathcal{O}(1 \%)$, the oscillations associated with $\Delta m^2_{14}$ will not be discernible if $\Delta_{14} \lesssim 10^{-2}$ over the entire range of reconstructed neutrino energies. For long-baseline oscillation experiments with $L \sim \mathcal{O}(10^3 \text{ km})$ and $E_\nu \sim \mathcal{O}(1-10 \text{ GeV})$, this condition translates into $\Delta m^2_{14} \lesssim 10^{-4}$ eV$^2$.\footnote{Note that $\Delta m^2_{12}$ is close to this limit, i.e., the wavelengths of its associated oscillations are too long to significantly impact oscillations at such an experiment. Nonetheless, sensitivity to the oscillations associated with $\Delta m^2_{12}$ comes from the interference with the oscillations due to $\Delta m^2_{13}$. As we discuss in Appendix~A, long-baseline experiments rely on information regarding $\Delta m^2_{12}$ and $\theta_{12}$ from other sources, including solar and reactor neutrinos, in order to precisely measure all oscillation parameters.}
Nonetheless, in this scenario, oscillations can be distinct from those among only three neutrinos. Here, the elements $U_{\alpha i}$, $\alpha=e,\mu,\tau$; $i=1,2,3$ do not form a unitary matrix and the number of independent parameters is larger than four, including sources of $CP$-invariance violation beyond the phase $\eta_{1}$ \cite{Hollander:2014iha,Klop:2014ima}. We return to this in Sec.~\ref{DUNE}.

When $\Delta_{14} \gg 1$ but $\nu_4$ is light enough to be produced coherently in the initial neutrino state, the oscillations of $\Delta m^2_{14}$ can be too rapid to be resolved by the finite energy resolution employed by the experiment. The oscillations of $\Delta m^2_{14}$ average out if $\Delta_{14}\times (\delta E/E_\nu)$ is, roughly,  larger than $2\pi$, where the energy bin width is $\delta E$ and the bin's central energy is $E_\nu$. For $L \sim \mathcal{O}(10^3 \text{ km})$, $E_\nu \sim \mathcal{O}(1-10 \text{ GeV})$, and $\delta E \sim$ 0.25 GeV, this will occur if $\Delta m^2_{14} \gtrsim 1$ eV$^2$. In this case, 
\begin{equation}
\label{prob_ave}
P_{\alpha\beta} \simeq \bigg\vert  \delta_{\alpha\beta} - U_{\alpha4}U_{\beta4}^* + U_{\alpha 2}U_{\beta 2}^* \left( e^{-i \Delta_{12}}-1 \right) + U_{\alpha 3}U_{\beta 3}^* \left( e^{-i \Delta_{13}}-1 \right) \bigg\vert^2 + \left|U_{\alpha4}U^*_{\beta4}\right|^2.
\end{equation}
This limit is, as far as measurements of the oscillation probabilities are concerned, equivalent to the decoherence of $\nu_4$ from the other neutrinos. As in the $\Delta_{14}\ll 1$ limit, oscillations are in general distinct from those among only three neutrinos. For example, one is, in principle, also sensitive to sources of $CP$-invariance violation beyond the phase $\eta_{1}$ \cite{Hollander:2014iha,Klop:2014ima}. 

When neutrinos propagate through matter, elastic, coherent, forward scattering modifies the oscillation probabilities in a well-known way. This can be parameterized via an effective potential generated by the background of electrons, protons and neutrons. The Hamiltonian $\delta H_{\alpha\beta}$ that describes neutrino oscillations, in the flavor basis, is \cite{Albright:2000xi}
\begin{equation}
\label{matterpotential}
\left( \frac{\delta H_{\alpha\beta}}{1 \text{ km}^{-1}} \right) = \left(\frac{A}{\text{1 eV}^2}\right) \delta_{\alpha e}\delta_{\beta e} + \left(\frac{A^\prime}{\text{1 eV}^2}\right) \delta_{\alpha s}\delta_{\beta s},
\end{equation}
where $(A/\text{1 eV}^2) = (3.85 \times 10^{-4}) Y_e (\rho/\text{1 g cm}^{-3})$ characterizes the charged-current interactions, $(A^\prime/\text{1 eV}^2) = (1.92 \times 10^{-4}) (1 - Y_e) (\rho/\text{1 g cm}^{-3})$ characterizes the neutral-current interactions, $Y_e$ is the electron fraction for the matter background, and $\rho$ is the density of the background.\footnote{The zero-point of the potential has been shifted in Eq.~(\ref{matterpotential}) so that the neutral-current contribution appears with the opposite sign in the sterile--sterile part of the Hamiltonian.}
The signs of $A$ and $A^\prime$ are flipped for antineutrinos. The Earth's crust typically has $Y_e \simeq 0.5$ and $\rho \simeq 3 \text{ g cm}^{-3}$~\cite{Albright:2000xi}. In the presence of matter, the Hamiltonian is no longer diagonal in the mass basis and the exact expressions for the oscillation probabilities are much more cumbersome. In our analyses, we treat the flavor evolution of the neutrino states numerically.

Our analysis makes use of initially muon-type neutrinos produced in pion decay to study $P_{\mu\mu}$ and $P_{\mu e}$. Because of experimental challenges involved in working with $\tau$ leptons, we do not consider oscillations into $\nu_\tau$.\footnote{The study of tau appearance requires neutrino energies above the tau-production threshold for neutrino--nucleon scattering, around 3.4~GeV. Hence, for the energies under consideration here, tau-appearence is severely phase-space suppressed. Furthermore, detectors must be able to identify taus with nonzero efficiency, an issue that is actively under investigation.}  Consequently, we do not expect to learn much about $\phi_{34}$, which only appears in the matrix elements $U_{\tau i}$. For long-baseline oscillation experiments with $L \sim \mathcal{O}(10^3 \text{ km})$ and $E_\nu \sim \mathcal{O}(1-10 \text{ GeV})$, $P_{\mu\mu}$ is mostly sensitivity to $\phi_{24}$, while $P_{\mu e}$ is sensitive to both $\phi_{24}$ and $\phi_{14}$, mostly via the product $\sin\phi_{24}\sin\phi_{14}$. Therefore, we expect DUNE to have greater sensitivity to $\phi_{24}$ than to $\phi_{14}$. Furthermore, these two channels both depend on the $CP$-odd phase $\eta_1$, as well as the combination 
\begin{equation}
\eta_s \equiv \eta_2 - \eta_3.
\end{equation}
In order to distinguish the effects of $\eta_2$ from $\eta_3$, one requires information regarding $U_{\tau i}$, which, as just argued above, is unavailable in the absence of searches for $\tau$ appearance or disappearance.

%%%%%

\setcounter{footnote}{0}
\setcounter{equation}{0}
\section{Experimental Sensitivity to a fourth neutrino at DUNE}
\label{DUNE}

We investigate the sensitivity of the proposed Deep Underground Neutrino Experiment (DUNE)~\cite{Adams:2013qkq} to a fourth neutrino. We consider that DUNE consists of a $34$ kiloton liquid argon detector and utilizes $1.2$~MW neutrino and antineutrino beams originating $1300$ km upstream at Fermilab, consistent with the proposal in Ref.~\cite{Adams:2013qkq}. We also simulate that the detector has resolution of $\sigma$ [GeV]$= 0.15/\sqrt{E_\nu\ [\text{GeV}]}$ for identifying electrons and $\sigma$ [GeV]$= 0.20/\sqrt{E_\nu\ [\text{GeV}]}$ for identifying muons. The neutrino energy ranges between $0.5$ and $20$ GeV and the flux is largest around $3.0$ GeV. In the following analyses, we simulate six years of data collection: 3 years each with the neutrino and antineutrino beams. 

\begin{figure}[htbp]
	\begin{center}
	\subfigure[]{\label{4Pane_Nu_App}\includegraphics[width=0.45\linewidth]{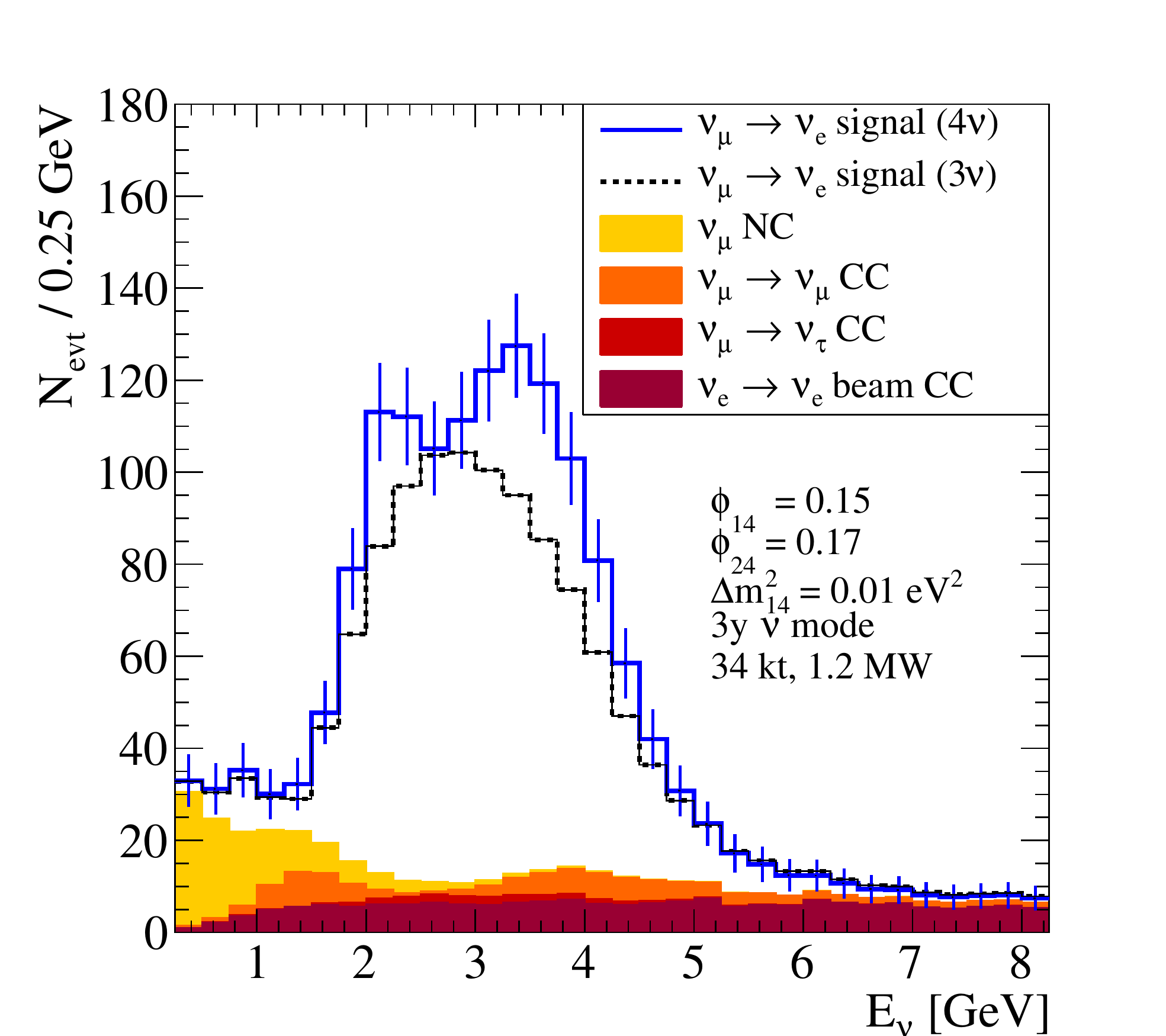}}
	\subfigure[]{\label{4Pane_NuBar_App}\includegraphics[width=0.45\linewidth]{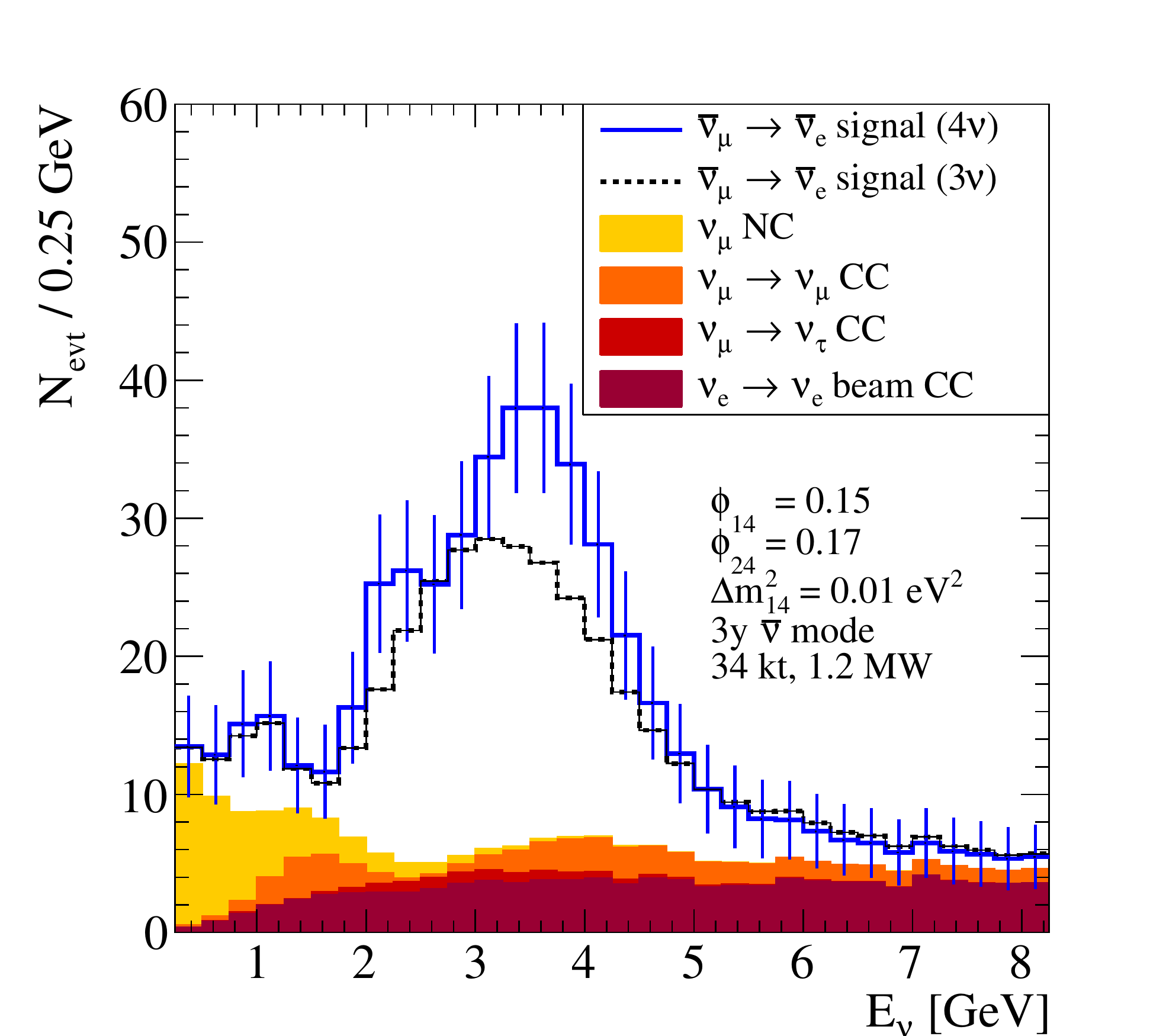}}
	\hspace{0.5in}
	\subfigure[]{\label{4Pane_Nu_Dis}\includegraphics[width=0.45\linewidth]{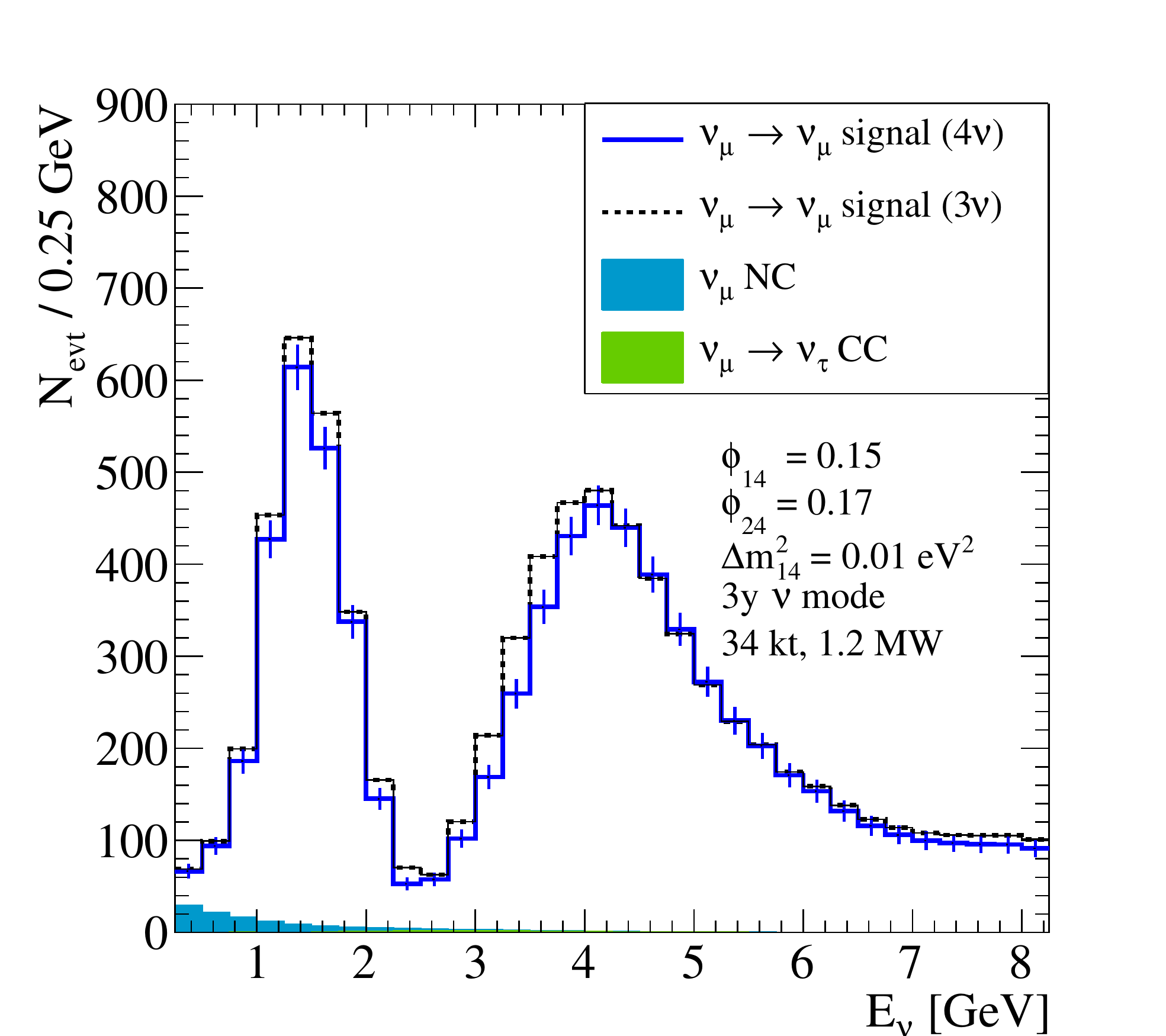}}
	\subfigure[]{\label{4Pane_NuBar_Dis}\includegraphics[width=0.45\linewidth]{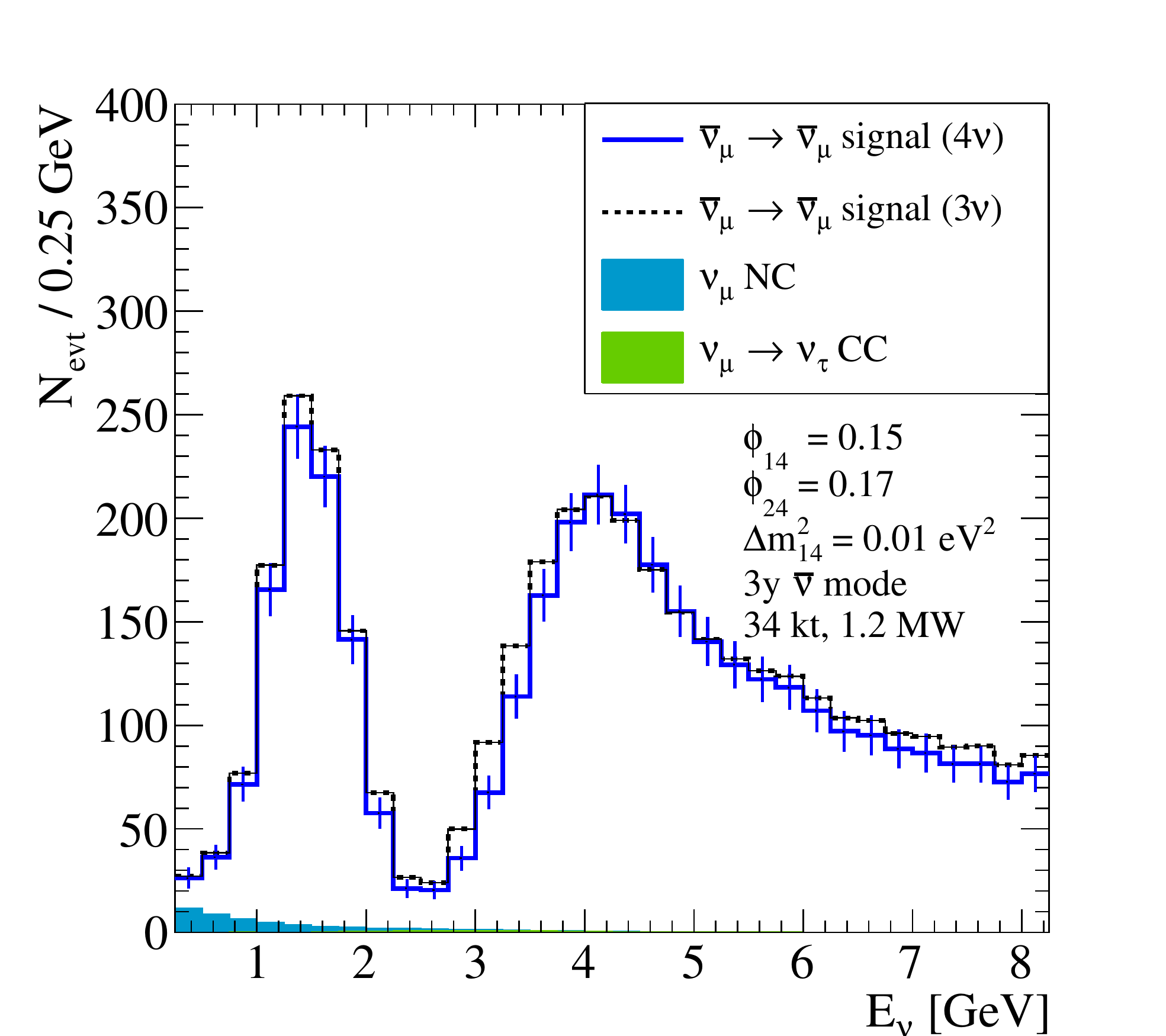}}
	\caption{Expected signal and background yields for six years (3y $\nu\ +$ 3y $\overline{\nu}$) of data collection at DUNE, using fluxes projected by Ref.~\cite{Adams:2013qkq}, for a 34 kiloton detector, and a 1.2 MW beam. (a) and (b) show appearance channel yields for neutrino and antineutrino beams, respectively, while (c) and (d) show disappearance channel yields. The $3\nu$ signal corresponds to the standard three-neutrino hypothesis, where $\sin^2\theta_{12} = 0.308,\ \sin^2\theta_{13} = 0.0235,\ \sin^2\theta_{23} = 0.437,\ \Delta m_{12}^2 = 7.54 \times 10^{-5}$ eV$^2$, $\Delta m_{13}^2 = 2.43\times 10^{-3}$ eV$^2$, $\delta_{CP} = 0$, while the $4\nu$ signal corresponds to $\sin^2\phi_{12} = 0.315$, $\sin^2\phi_{13} = 0.024$, $\sin^2\phi_{23} = 0.456$, $\sin^2\phi_{14} = 0.023$, $\sin^2\phi_{24} = 0.030$, $\Delta m_{14}^2 = 10^{-2}$ eV$^2$, $\eta_{1} = 0$, and $\eta_s = 0$. Statistical uncertainties are shown as vertical bars in each bin. Backgrounds are defined in the text and are assumed to be identical for the three- and four-neutrino scenarios: any discrepancy is negligible after accounting for a 5\% normalization uncertainty.}
	\label{fig:4Pane}
	\end{center}
\end{figure}

We use the neutrino fluxes and signal reconstruction efficiencies projected by DUNE (Fig.~3.18 and Table 4.2 in Ref.~\cite{Adams:2013qkq}, respectively) and the neutrino--nucleon cross-sections reported in Ref.~\cite{Formaggio:2013kya} to calculate expected yields. For a three-neutrino scenario, we use input values consistent with the best-fit results compiled in Ref.~\cite{Agashe:2014kda}: $\sin^2\theta_{12} = 0.308$, $\sin^2\theta_{13} = 0.0235$, $\sin^2\theta_{23} = 0.437$, $\Delta m_{12}^2 = 7.54\times 10^{-5}$ eV$^2$, $\Delta m_{13}^2 = +2.43\times 10^{-3}$ eV$^2$ (hence a normal neutrino mass hierarchy), and $\delta_{CP} = 0$. The four dominant backgrounds are consequences of muon-type neutrino neutral-current scattering (``$\nu_\mu$ NC''), tau-type neutrino charged-current scattering (``$\nu_\mu \rightarrow \nu_\tau$ CC''), muon-type neutrino charged-current scattering (``$\nu_\mu \rightarrow \nu_\mu$ CC''), and beam electron-type neutrino charged-current scattering (``$\nu_e\rightarrow \nu_e$ beam CC''), depicted in Figs.~\ref{fig:4Pane}(a)-(d). The rates associated with these backgrounds are taken from Ref.~\cite{Adams:2013qkq}. We reproduce the signal and background yields in Ref.~\cite{Adams:2013qkq} for the appearance ($P_{\mu e}$) and disappearance ($P_{\mu\mu}$) channels, shown as dashed lines in Figs.~\ref{fig:4Pane}(a)-(d), i.e., ``$\nu_\mu \rightarrow \nu_e$ signal 3$\nu$'' and ``$\nu_\mu \rightarrow \nu_\mu$ signal 3$\nu$,'' respectively. In Appendix~A, we demonstrate comparable sensitivity to those computed in Ref.~\cite{Adams:2013qkq}.

In order to illustrate the effects of a fourth neutrino, the expected yields along with the three-neutrino yields in Fig.~\ref{fig:4Pane} are depicted for $\sin^2\phi_{14} = 0.023$, $\sin^2\phi_{24} = 0.030$, $\sin^2\phi_{34} = 0$, $\Delta m_{14}^2 = 10^{-2}$ eV$^2$, and $\eta_s = 0$ (``$\nu_\mu \rightarrow \nu_e$ signal 4$\nu$'' and ``$\nu_\mu \rightarrow \nu_\mu$ signal 4$\nu$''). We choose the value of $\Delta m_{14}^2$ such that several oscillations due to the fourth neutrino occur within the energy window of the experiment. Here, the input values of $\phi_{12}$, $\phi_{13}$, and $\phi_{23}$ are slightly different from the values mentioned above for $\theta_{12}$, $\theta_{13}$, and $\theta_{23}$, and are chosen so that the values of $|U_{e2}|^2$, $|U_{e3}|^2$, and $|U_{\mu 3}|^2$ are consistent with three-flavor fits to the neutrino data~\cite{Agashe:2014kda}.\footnote{We do not have the freedom to set all nine matrix elements $U_{\alpha i}$ ($\alpha = e$, $\mu$, $\tau$; $i = 1$, $2$, $3$) equal to their three-neutrino best-fit values. Explicitly, we choose $|U_{e2}|^2 = 0.301$, $|U_{e3}|^2 = 0.023$, and $|U_{\mu 3}|^2 = 0.427$.}

\subsection{Constraining the Four-Neutrino Hypothesis}
\label{subsec:exclude}
\begin{figure}[htbp]
	\begin{center}
	\subfigure[]{\label{subfig:T14xDm14}\includegraphics[width=0.49\linewidth]{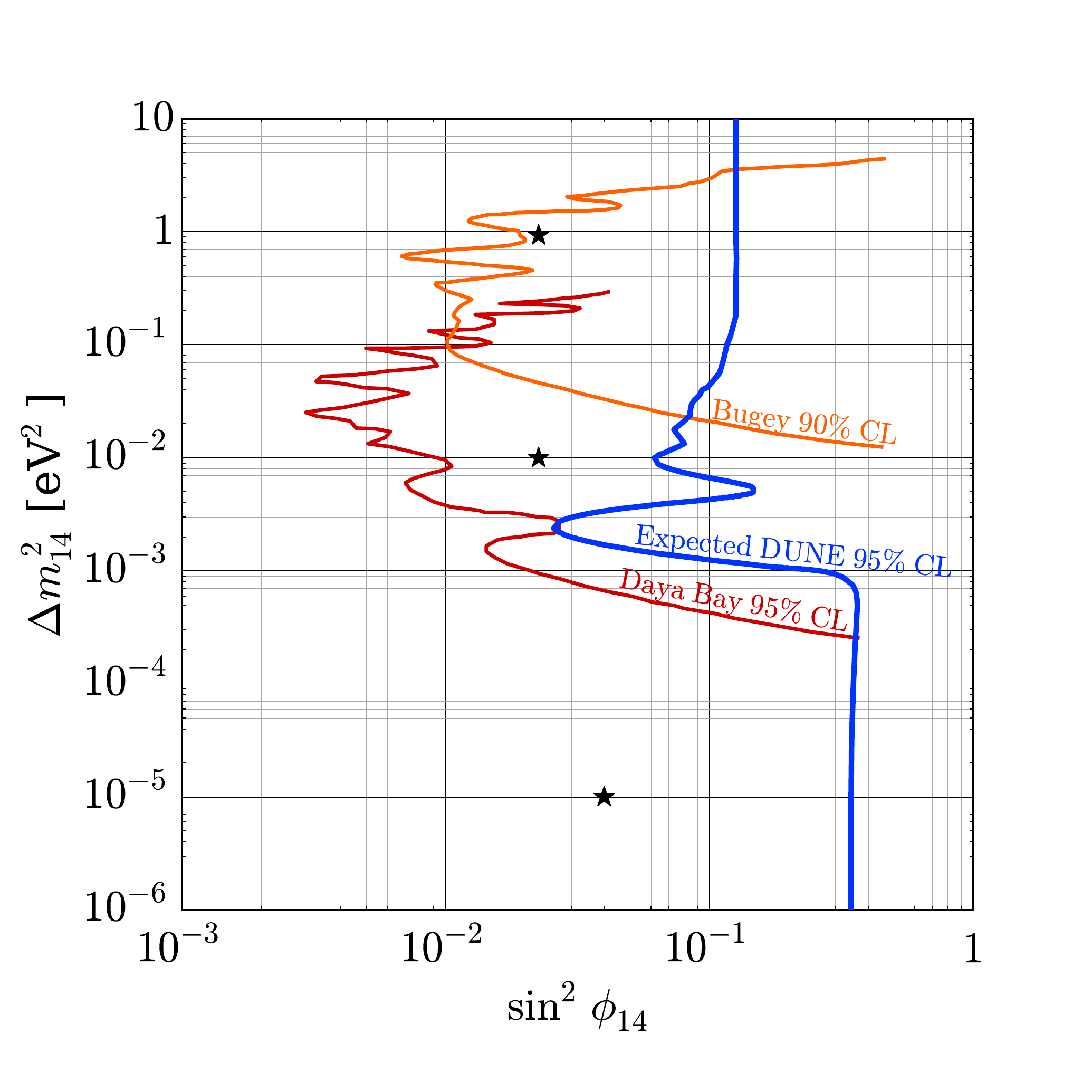}}
	\subfigure[]{\label{subfig:T24xDm14}\includegraphics[width=0.49\linewidth]{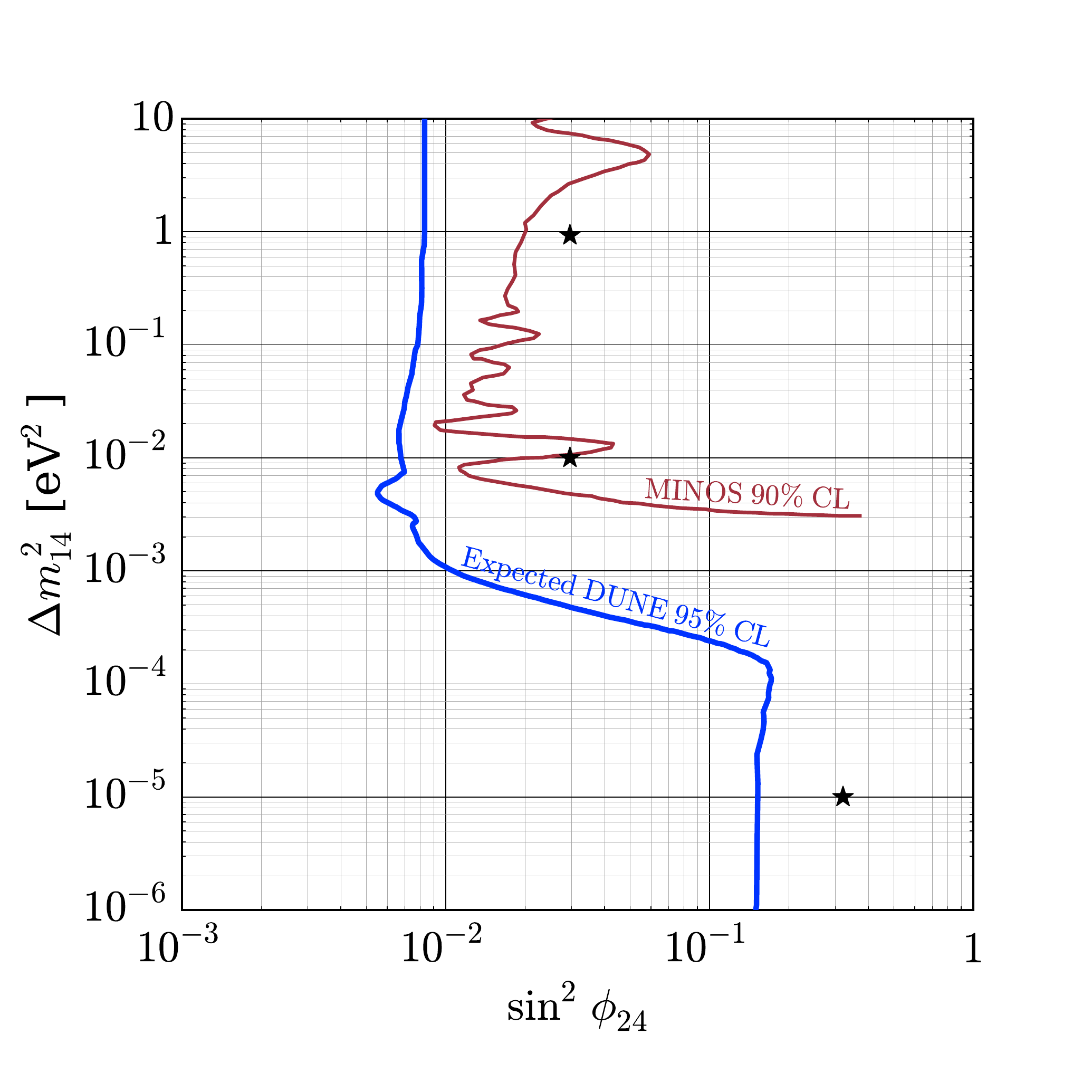}}
	\caption{Expected exclusion limits for $\sin^2{\phi_{14}}$ (a) and $\sin^2{\phi_{24}}$ (b) vs.\ $\Delta m_{14}^2$ at DUNE (blue), assuming a $34$ kiloton detector and a $1.2$	MW beam with six years (3y $\nu\ +$ 3y $\overline{\nu}$) of data collection. The exclusion limits become independent of $\Delta m_{14}^2$ when the mass-squared difference is large ($\gtrsim 10^{-1}$ eV$^2$) or small ($\lesssim 10^{-4}$ eV$^2$). Results from the Daya Bay~\cite{An:2014bik} (red) and Bugey ~\cite{Declais:1994su} (orange) are shown in (a), and results from MINOS~\cite{Timmons:2015lga} (maroon) are shown in (b). The three sets of four-neutrino parameters we consider in Section~\ref{subsec:MeasFourth}, listed in Table~\ref{CaseTable}, are denoted by black stars above.}
	\label{fig:MassSensitivity}
	\end{center}
\end{figure}

If the data are consistent with the three-neutrino hypothesis -- the three-neutrino scenario outlined above -- one can place upper bounds on the values of $\phi_{14}$ and $\phi_{24}$ for given values of $\Delta m^2_{14}$. We calculate 95\% confidence level (CL) exclusion limits for a fourth neutrino in the $\sin^2{\phi_{14}}$ - $\Delta m_{14}^2$ and $\sin^2{\phi_{24}}$ - $\Delta m_{14}^2$ planes, depicted in Fig.~\ref{fig:MassSensitivity}, using the appearance and disappearance channels and assuming running for three years each with the neutrino and antineutrino beams. We include normalization uncertainties of 1\% and 5\% for the signal and background yields, respectively.

Fig.~\ref{fig:MassSensitivity}(a) also depicts the results from the Daya Bay~\cite{An:2014bik} and Bugey~\cite{Declais:1994su} experiments in the $\sin^2{\phi_{14}}$ - $\Delta m_{14}^2$ plane. The existing experiments have greater sensitivity to $\sin^2{\phi_{14}}$ for values of $\Delta m^2_{14}\gtrsim 10^{-4}$~eV$^2$. For smaller values of $\Delta m^2_{14}\lesssim 10^{-4}$~eV$^2$, none of the experimental probes, including DUNE, can ``see'' the very long new oscillation length. Nonetheless, since DUNE measures both appearance and disappearance, we can constrain very large values of $\sin^2{\phi_{14}}$ as these render the upper-left $3\times 3$ mixing sub-matrix unacceptably non-unitary. The same phenomenon can be observed in the  $\sin^2{\phi_{24}}$ - $\Delta m_{14}^2$ plane (Fig.~\ref{fig:MassSensitivity}(b)). Very large values of  $\sin^2{\phi_{24}}$ are ruled out, even for very small values of $\Delta m^2_{14}$.

In the $\sin^2{\phi_{24}}$ - $\Delta m_{14}^2$ plane (Fig.~\ref{fig:MassSensitivity}(b)), we also show results from the MINOS~\cite{Timmons:2015lga} experiment and note that DUNE will be sensitive to lower values of the mixing angle and the mass-squared difference, due to DUNE having greater expected yield and a broader range of $L/E_{\nu}$ values. Because the disappearance channels depend strongly on $|U_{\mu 4}|^2$, and have higher yields than the appearance channels, DUNE has greater sensitivity to $\phi_{24}$ than $\phi_{14}$. We also note that, as expected and discussed in the previous section, if the mass-squared difference is either very small, $\Delta m_{14}^2 \lesssim 10^{-4}$ eV$^2$, or very large, $\Delta m_{14}^2 \gtrsim 1$ eV$^2$, the limits are independent of the new mass-squared difference. These ranges correspond, respectively, to $\Delta_{14} \lesssim 10^{-2}$, where oscillations due to the fourth neutrino are undetectable over the energy range of the experiment, and to the oscillations associated to the new frequency averaging out over the width of the energy bin at DUNE (here $\Delta E_\nu$, the width of a bin, is $0.25$ GeV). Finally, the sensitivity in the $\sin^2\phi_{24}$ - $\Delta m_{14}^2$ plane extends to lower values of $\Delta m_{14}^2$ than that in the $\sin^2{\phi_{24}}$ - $\Delta m_{14}^2$ plane due to the higher yield of the disappearance channel.

\begin{figure}[htbp]
	\begin{center}
	\includegraphics[width=0.6\linewidth]{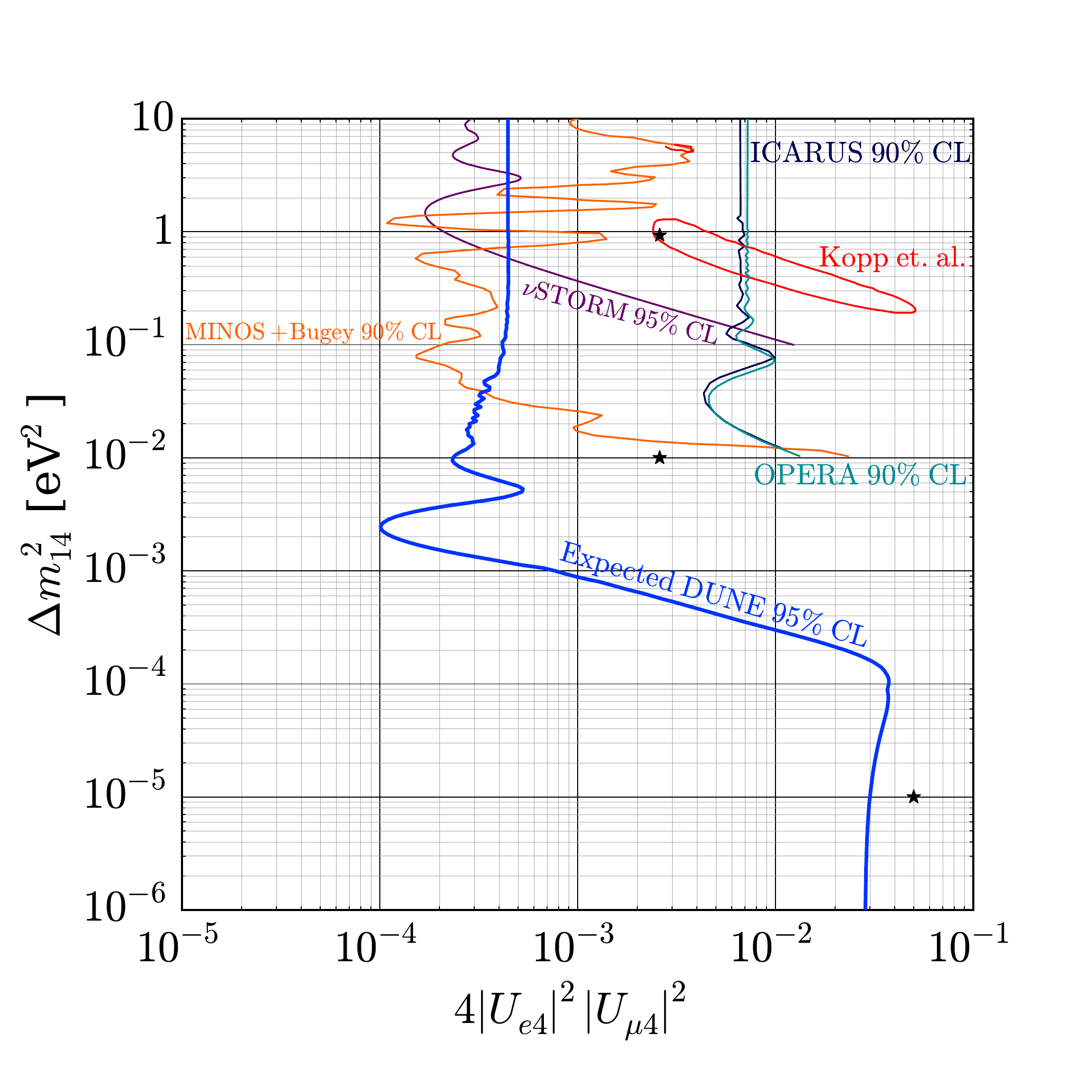}
	\caption{Exclusion limits in the $4|U_{e4}|^2|U_{\mu 4}|^4$ - $\Delta m_{14}^2$ plane for various existing and proposed neutrino experiments. Expected exclusion limits are shown at 95\% CL for the proposed DUNE (blue) and $\nu$STORM~\citep{Kyberd:2012iz} (purple) experiments. Results are shown at 90\% CL for the MINOS and Bugey~\citep{Timmons:2015lga} (orange), OPERA~\citep{Agafonova:2013xsk} (teal), and the ICARUS~\citep{Antonello:2013gut} (dark blue) experiments. Additionally, the fit to the $3+1$ scenario including the short-baseline anomalies, reported in Ref.~\cite{Kopp:2013vaa}, is shown. The three sets of four-neutrino parameters we consider in Section~\ref{subsec:MeasFourth}, listed in Table~\ref{CaseTable}, are denoted by black stars above.}
	\label{SBLComp}
	\end{center}
\end{figure}

Additionally, we calculate exclusion limits at 95\% CL in the $4|U_{e4}|^2 |U_{\mu 4}|^2$ - $\Delta m_{14}^2$ 
plane\footnote{In the limit when oscillations due to the new mass-squared difference are dominant, the appearance channel oscillation probability takes the simple form
\begin{align}
P_{\mu e} \simeq 4|U_{e4}|^2|U_{\mu 4}|^2 \sin^2\left({\frac{\Delta m_{14}^2 L}{4 E_\nu}}\right) \equiv \sin^2{\left(2\theta_{e\mu}\right)}\sin^2\left({\frac{\Delta m_{14}^2 L}{4 E_\nu}}\right).
\end{align}
The effective mixing angle $\theta_{e\mu}$ is commonly used in the literature for $\nu_\mu \rightarrow \nu_e$ short-baseline appearance searches~(see, for example, \citep{Adey:2014rfv, deGouvea:2014aoa,Timmons:2015lga,Agafonova:2013xsk,Antonello:2013gut}).} 
in order to compare the DUNE sensitivity to the proposed short-baseline experiment $\nu$STORM~\cite{Kyberd:2012iz}, and the current long-baseline experiments MINOS, OPERA, and ICARUS~\citep{Timmons:2015lga,Agafonova:2013xsk,Antonello:2013gut}. Additionally, we include the results of the global fit to all neutrino data assuming a four-neutrino scenario, reported in Ref.~\cite{Kopp:2013vaa}. This fit includes data from short-baseline experiments, including the short-baseline anomalies discussed earlier. In Fig.~\ref{SBLComp}, we see that DUNE is sensitive to lower values of $\Delta m_{14}^2$ than any existing or proposed experiment due to its access to a wider range of $L/E_\nu$ values.

\subsection{Measuring the New Mixing Parameters}
\label{subsec:MeasFourth}
Assuming the existence of a fourth neutrino, we explore the capability of DUNE to measure the new mixing angles and mass-squared difference. We choose three sets of parameters, listed in Table~\ref{CaseTable} and denoted by black stars in Figs.~\ref{fig:MassSensitivity} and \ref{SBLComp}, and calculate expected yields for the appearance and disappearance channels. Case 1 is consistent with the global four-neutrino fit performed in Ref.~\cite{Kopp:2013vaa} (the red ellipse in Fig.~\ref{SBLComp}). Here, $\Delta m^2_{14}$ is large enough that we expect the oscillations associated to the new (very short) oscillation length to average out at DUNE.  Case 2 uses the same mixing angles as Case 1, but with a lower value of $\Delta m_{14}^2$. The parameters are within the the reach of DUNE, but outside the reach of current and proposed short-baseline experiments.\footnote{Note, however, that Case 2 is in slight disagreement with existing bounds from Daya Bay and MINOS, see Fig.~\ref{fig:MassSensitivity}.} Here, $\Delta m^2_{14}$ is small enough that we expect the oscillations associated to the new oscillation length to be visible at DUNE. Case 3 has a much lower mass-squared difference, $\Delta m_{14}^2 = 10^{-5}$ eV$^2$, but has large values of $\phi_{14}$ and $\phi_{24}$.\footnote{A recent analysis of solar and reactor data constrain $\sin^2\phi_{14}\lesssim 0.04$~\cite{Palazzo:2012yf}. This bound is not depicted in Fig.~\ref{fig:MassSensitivity}.} Here, $\Delta m^2_{14}$ is too small to be seen at DUNE. Nonetheless, as discussed earlier, the new mixing angles are large enough that nontrivial information on $\phi_{14}$ and $\phi_{24}$ can be extracted.
\begin{table}[htbp]
\centering
\begin{tabular}{| c || c  | c | c | c || c | c | c | c | c | c | }
\hline
 & $\sin^2\phi_{14}$ & $\sin^2\phi_{24}$ & $\Delta m_{14}^2$ (eV$^2$) & $\eta_s$ & $\sin^2\phi_{12}$ & $\sin^2\phi_{13}$ & $\sin^2\phi_{23}$ & $\Delta m_{12}^2$ (eV$^2$) & 
 $\Delta m_{13}^2$ (eV$^2$) & $\eta_{1}$ \\
 \hline
 \textbf{Case 1} & 0.023 & 0.030 & 0.93 & $-\pi/4$ & 0.315 & 0.0238 & 0.456 & $7.54\times 10^{-5}$ & $2.43\times 10^{-3}$ & $\pi/3$\\
 \hline
 \textbf{Case 2} & 0.023 & 0.030 & $1.0\times 10^{-2}$ & $-\pi/4$ & 0.315 & 0.0238 & 0.456 & $7.54\times 10^{-5}$ & $2.43\times 10^{-3}$  & $\pi/3$\\
 \hline
 \textbf{Case 3} & 0.040 & 0.320 & $1.0\times 10^{-5}$ & $-\pi/4$ & 0.321 & 0.0244 & 0.639 & $7.54\times 10^{-5}$ & $2.43\times 10^{-3}$ & $\pi/3$\\
 \hline
\end{tabular}
\caption{Input values of the parameters for the three scenarios considered for the four-neutrino hypothesis. Values of $\phi_{12}$, $\phi_{13}$, and $\phi_{23}$ are chosen to be consistent with the best-fit values of $|U_{e2}|^2$, $|U_{e3}|^2$, and $|U_{\mu 3}|^2$, given choices of $\phi_{14}$ and $\phi_{24}$. Here, $\eta_s \equiv \eta_2 - \eta_3$. Note that $\Delta m^2_{14}$ is explicitly assumed to be positive, i.e., $m_4^2>m_1^2$.}
\label{CaseTable}
\end{table}

\begin{figure}[htbp]
    \begin{center}
    \includegraphics[width=1\textwidth]{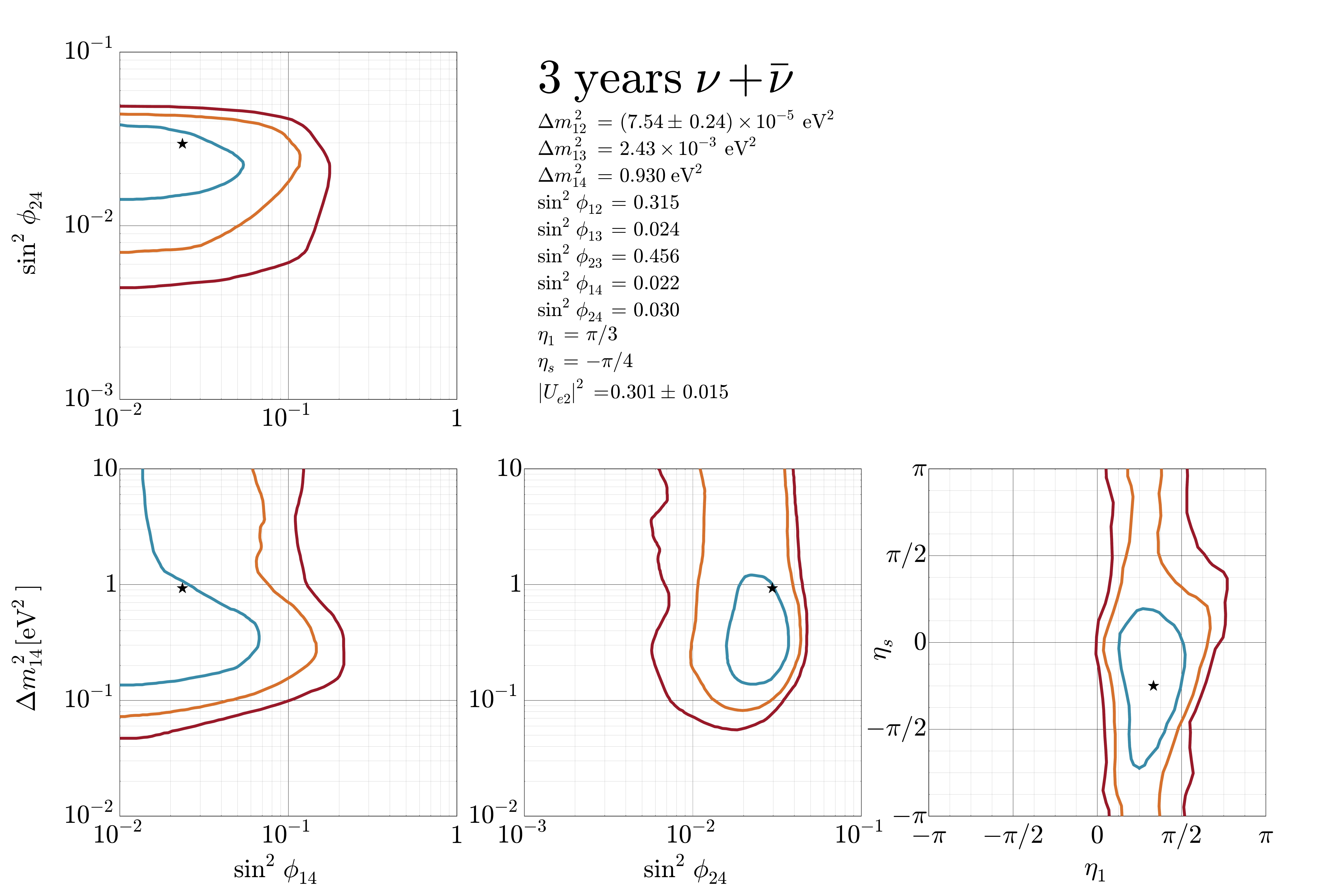}
    \caption{Expected sensitivity contours at 68.3\% (blue), 95\% (orange), and 99\% (red) CL at DUNE with six years of data collection (3y $\nu\ +$ 3y $\overline{\nu}$), a 34 kiloton detector, and a 1.2 MW beam given the existence of a fourth neutrino with parameters from Case~1 in Table~\ref{CaseTable}. Results from solar neutrino experiments are included here as Gaussian priors for the values of $|U_{e2}|^2 = 0.301 \pm 0.015$ and $\Delta m_{12}^2 = 7.54\pm 0.24\times 10^{-5}$ eV$^2$~\citep{Agashe:2014kda}. }
    \label{fig:DCPPi3_DSNPi4_KoppAngles_KoppMass}
    \end{center}
\end{figure}

\begin{figure}[htbp]
    \begin{center}
   \includegraphics[width=1\textwidth]{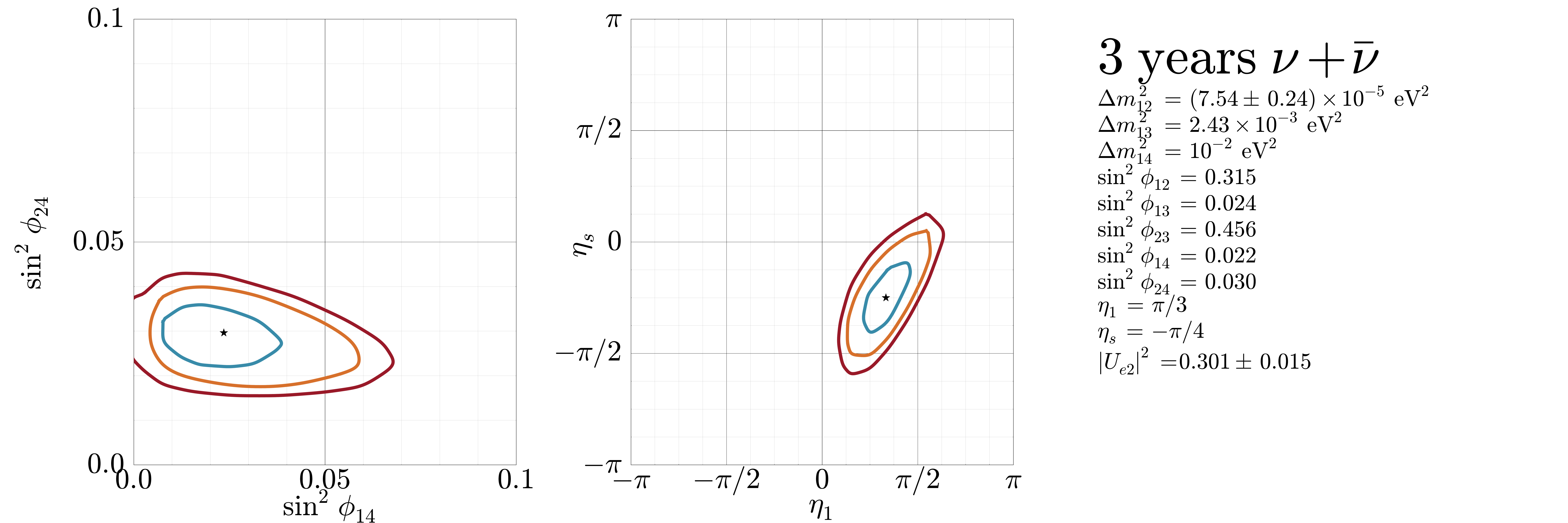}
    \caption{Expected sensitivity contours at 68.3\% (blue), 95\% (orange), and 99\% (red) CL at DUNE with six years of data collection (3y $\nu\ +$ 3y $\overline{\nu}$), a 34 kiloton detector, and a 1.2 MW beam given the existence of a fourth neutrino with parameters from Case~2 in Table~\ref{CaseTable}. Results from solar neutrino experiments are included here as Gaussian priors for the values of $|U_{e2}|^2 = 0.301 \pm 0.015$ and $\Delta m_{12}^2 = 7.54\pm 0.24\times 10^{-5}$ eV$^2$~\citep{Agashe:2014kda}. }
    \label{fig:DCPPi3_DSNPi4_KoppAngles_IntMass}
    \end{center}
\end{figure}

\begin{figure}[htbp]
	\begin{center}
	\includegraphics[width=0.7\linewidth]{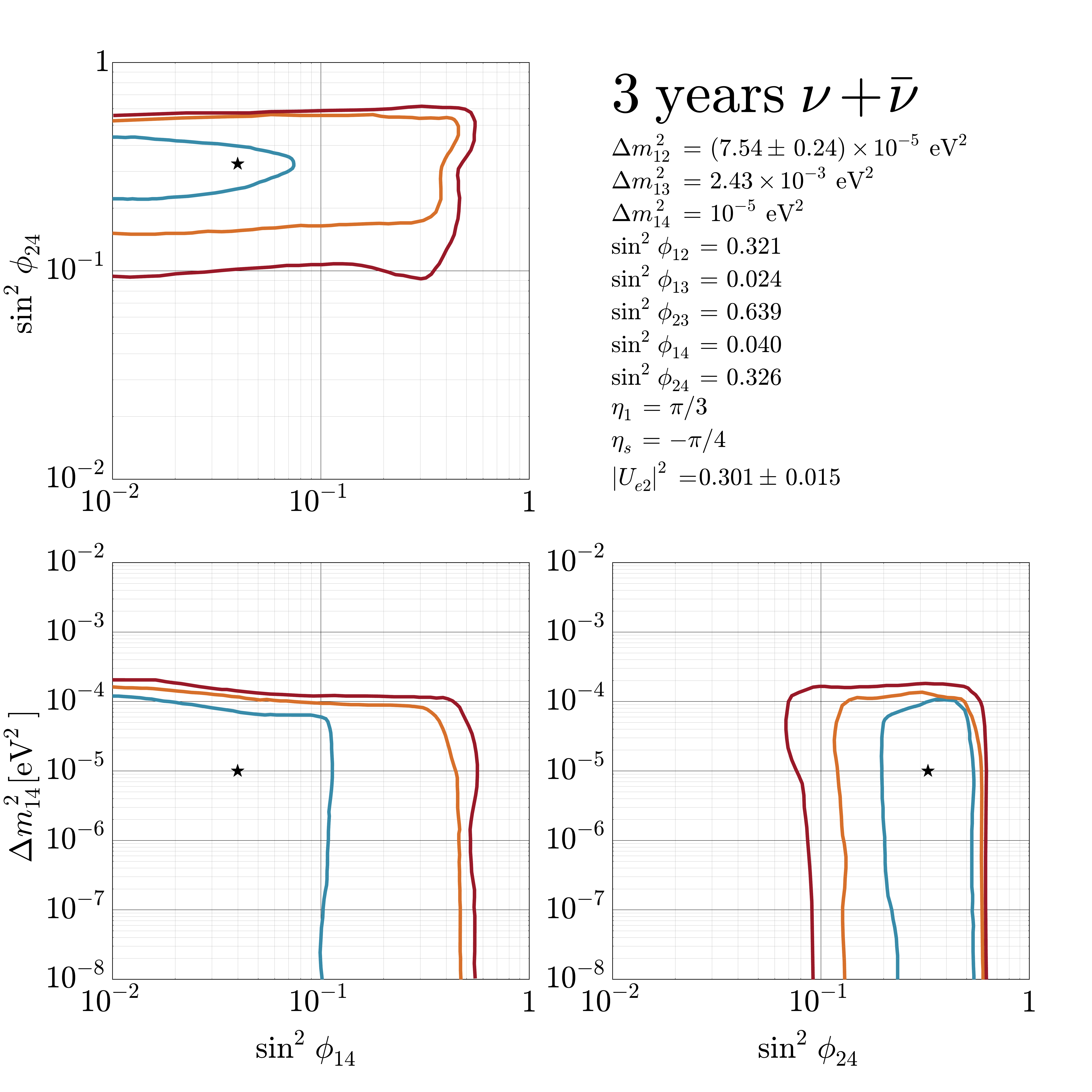}
	\caption{Expected sensitivity contours at 68.3\% (blue), 95\% (orange), and 99\% (red) CL at DUNE with six years of data collection (3y $\nu\ +$ 3y $\overline{\nu}$), a 34 kiloton detector, and a 1.2 MW beam given the existence of a fourth neutrino with parameters from Case~3 in Table~\ref{CaseTable}. Results from solar neutrino experiments are included here as Gaussian priors for the values of  $|U_{e2}|^2 = 0.301 \pm 0.015$ and $\Delta m_{12}^2 = 7.54\pm 0.24\times 10^{-5}$ eV$^2$~\citep{Agashe:2014kda}. }
	\label{fig:DCPPi3_DSNPi4_NewAngles_LowMass}
	\end{center}
\end{figure}

In all three Cases, we assume that the neutrino mass hierarchy for the mostly active states is normal, i.e.\ $\Delta m_{13}^2 = +2.43\times 10^{-3}$ eV$^2$, and in all Cases we assume $\eta_{1} = \pi/3$ and $\eta_s = -\pi/4$, typical of scenarios where $CP$-invariance violating effects are large.\footnote{We explored several other sets of input values for $\eta_1$ and $\eta_s$. This particular choice leads to generically large effects without extraordinary cancellations, enhancements, or ambiguities.} For completeness, we also assume, in all Cases, $\sin^2\phi_{34}=0$. Gaussian priors are adopted, mostly from solar neutrino data and data from KamLAND, on the solar parameters, $|U_{e2}|^2 = 0.301 \pm 0.015$ and $\Delta m_{12}^2 = (7.54 \pm 0.24) \times 10^{-5}$ eV$^2$~\cite{Agashe:2014kda}. Without these priors, DUNE is mostly insensitive to either $\Delta m_{12}^2$ or $\phi_{12}$. We make use of the Markov Chain Monte Carlo package {\sc emcee}~\cite{ForemanMackey:2012ig}, which estimates a probability distribution for each fitting parameter. Figs.~\ref{fig:AppendixCase1}, \ref{fig:DCPPi3_DSNPi4_KoppAngles_1en2Mass}, and \ref{fig:AppendixCase3} in Appendix~B, depict sensitivity contours at 68.3\%, 95\%, and 99\% CL and one-dimensional $\chi^2$ distributions for the ten parameters, for Cases 1, 2, and 3, respectively. Input values from Table~\ref{CaseTable} are shown as stars in the two-dimensional plots. Given the fact that the amount of information in  Figs.~\ref{fig:AppendixCase1}, \ref{fig:DCPPi3_DSNPi4_KoppAngles_1en2Mass}, and \ref{fig:AppendixCase3} is somewhat overwhelming, in order to guide the following discussions, Figs.~\ref{fig:DCPPi3_DSNPi4_KoppAngles_KoppMass}, \ref{fig:DCPPi3_DSNPi4_KoppAngles_IntMass}, and \ref{fig:DCPPi3_DSNPi4_NewAngles_LowMass} depict sensitivity contours at 68.3\%, 95\%, and 99\% CL for a subset of the parameters of interest, for Cases 1, 2, and 3, respectively.

Fig.~\ref{fig:DCPPi3_DSNPi4_KoppAngles_KoppMass} depicts the fit results for a subset of the parameters ($\sin^2\phi_{24}$, $\sin^2\phi_{14}$, $\eta_1$, $\eta_s$ and $\Delta m^2_{14}$), assuming Case~1. Here, the values of $\sin^2\phi_{24}$ and $\Delta m_{14}^2$ can be excluded from $0$ at the 99\% CL, while the value of $\sin^2\phi_{14}$ is consistent with $0$ at 68.3\% and $\eta_s$ cannot be constrained at the 95\% CL. Nonetheless, the $CP$-odd phase $\eta_1$, which can be more or less trivially associated with the $CP$-odd phase $\delta_{CP}$ in the three-neutrino scenario, is constrained to be nonzero at the 99\% CL. As expected, there is very little sensitivity to $\Delta m_{14}^2$, except for establishing that it is large ($\Delta m_{14}^2>7.9\times 10^{-2}$~eV$^2$ at the 99\% CL). 

Fig.~\ref{fig:DCPPi3_DSNPi4_KoppAngles_IntMass} depicts the fit results for a subset of the parameters ($\sin^2\phi_{24}$, $\sin^2\phi_{14}$, $\eta_1$, and $\eta_s$), assuming Case~2. Here, the values of $\sin^2\phi_{14}$, $\sin^2\phi_{24}$, $\Delta m_{14}^2$ (cf. Fig.~\ref{fig:DCPPi3_DSNPi4_KoppAngles_1en2Mass}), and $\eta_s$ are observed at at least the 95\% CL, i.e., the fit establishes that none of the new physics parameters vanish. In particular, the values of $\sin^2\phi_{14}$, $\sin^2\phi_{24}$, and $\Delta m_{14}^2$ are excluded from zero at the 99\% CL. In this case, there is enough sensitivity to the two independent $CP$-odd phases to establish that not only there are new neutrino degrees of freedom but that there is more than one new $CP$-invariance violating parameter in the theory. In summary, one can establish that there is new physics beyond the standard paradigm, and that the new physics is $CP$-invariance violating. 

Fig.~\ref{fig:DCPPi3_DSNPi4_NewAngles_LowMass} depicts the fit results for a subset of the parameters ($\sin^2\phi_{24}$, $\sin^2\phi_{14}$, and $\Delta m^2_{14}$), assuming Case~3. The results here are somewhat similar to (but less constraining than) those from Case 1. The measurement of $\Delta m_{14}^2$ is consistent with $0$ at 68.3\% CL. but, as expected, the data reveal that it is small ($\Delta m_{14}^2<1.6\times 10^{-4}$~eV$^2$ at the 99\% CL). The new $CP$-odd phase cannot be measured significantly (cf. Fig.~\ref{fig:AppendixCase3}). On the other hand, one can exclude the hypothesis that the ``standard'' $CP$-odd phase $\eta_1$ is zero, but the sensitivity is worse than what one can achieve if the data were consistent with the three-flavor scenario. 

If there is a fourth neutrino mass-eigenstate, the parameters of the fourth neutrino may significantly affect DUNE's ability to measure the mixing angles naively associated with three-neutrino oscillation. For example, as shown in Appendix~A and in Ref.~\cite{Adams:2013qkq}, the expected measurement precision for $\theta_{13}$ assuming a three-neutrino scenario is $\delta \theta_{13}/\theta_{13} \simeq 3\%$. In Cases 1 and 3, this precision is much worse, $\delta \phi_{13}/\phi_{13} \simeq 10\%$. In Case 2, however, the precision with which $\phi_{13}$ can be measured is $\delta \phi_{13}/\phi_{13} \simeq 4\%$, i.e., similar to the precision obtained in the three-neutrino scenario. This happens because, in Case 2, one can mostly disentangle effects due to the different oscillation frequencies. 

\subsection{Testing the Three-Massive-Neutrinos Paradigm}
\label{subsec:Incorrect}

In Sec.~\ref{subsec:exclude}, we simulated data assuming a three-neutrino scenario and, by analyzing it assuming the four-neutrino hypothesis, were able to constrain the values of the new mixing parameters. In Sec.~\ref{subsec:MeasFourth}, we simulated data assuming different four-neutrino scenarios and, by analyzing it assuming the four-neutrino hypothesis, were able to constrain or measure, sometimes quite precisely, the new mixing parameters. Here we address a different question: if we were to simulate data consistent with a four-neutrino scenario, would we be able to tell that there are more than three neutrinos? More concretely, would the analysis of the data assuming the three-massive-neutrinos paradigm reveal that the paradigm is incorrect? 

To address this question, we fit the expected yields from Case 2, introduced in Sec.~\ref{subsec:MeasFourth} (see Table~\ref{CaseTable}) assuming the three-neutrino hypothesis. We obtain best-fit values of $\theta_{12}$, $\theta_{13}$, $\theta_{23}$, $\Delta m_{12}^2$, $\Delta m_{13}^2$, and $\delta_{CP}$, along with associated uncertainties. The precision with which the parameters can be measured is comparable to what would be expected of DUNE if the data were consistent with the three-neutrino hypothesis. We also find, however, that the overall quality of the fit is poor: $\chi^2_{\text{min}} /$ degrees of freedom (dof) $\simeq 180/114$, or a discrepancy of roughly $4\sigma$. Hence, the three-neutrino hypothesis cannot mimic the additional oscillations associated with $\Delta m_{14}^2 = 10^{-2}$ eV$^2$, for any set of values of the three-neutrino parameters.\footnote{While we concentrate on Case 2 here, we obtain poor fits also by assuming data consistent with Cases 1 and 3, where the new oscillation frequency cannot be explicitly observed. The discrepancy is most significant for Case 2, however.}

\begin{figure}[htbp]
	\begin{center}
	\includegraphics[width=0.45\linewidth]{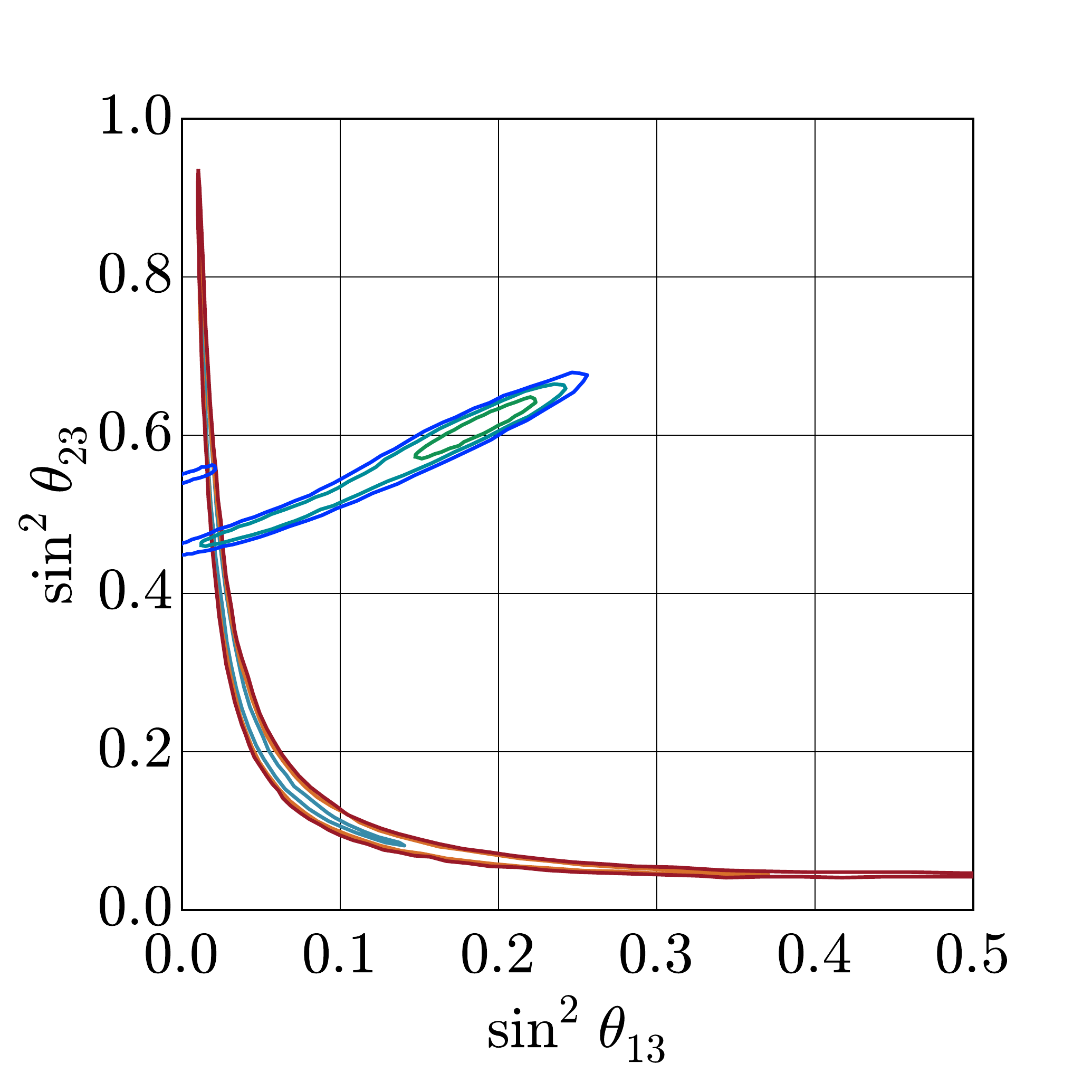}
	\caption{Expected sensitivity contours at 68.3\%, 95\%, and 99\% for neutrino and antineutrino appearance channels (blue, orange, red), and neutrino and antineutrino disappearance channels (green, teal, blue) in the $\sin^2{\theta_{13}}$ - $\sin^2{\theta_{23}}$ plane, assuming the data are consistent with Case~2 and analyzing it assuming the three-massive neutrinos paradigm.}
	\label{Case2:T13T23Comp}
	\end{center}
\end{figure}

Once a bad goodness-of-fit is established, it becomes crucial to identify in which way the three-neutrino hypothesis fails. This can be done in a variety of ways. Here, for illustrative purposes, we try to diagnose the poor goodness-of-fit by splitting the data set into two subsets: the appearance data and the disappearance data, and analyze both subsets separately (combining neutrino and antineutrino data in each case). In both sub-channels, the extraction of $\theta_{12}$ and $\Delta m_{12}^2$ is mostly driven by the priors from solar neutrino data, while the disappearance data are mostly insensitive to the $CP$-odd parameter $\delta_{CP}$. For these two reasons, it is most illuminating to examine the measurements obtained from these two fits in the $\sin^2\theta_{13}$ - $\sin^2\theta_{23}$ plane, depicted in Fig.~\ref{Case2:T13T23Comp}. Fig.~\ref{Case2:T13T23Comp} reveals that the appearance and disappearance channels favor different values of $\sin^2\theta_{13}$ and $\sin^2\theta_{23}$, with no overlap of the preferred regions at the 68.3\% CL. For the appearance channels, the fit has $\chi^2_{\text{min}}/$ dof $\simeq 78/54$ (roughly $2\sigma$), and for the disappearance channels, $\chi^2_{\text{min}}/$ dof $\simeq 91/54$ (roughly $3\sigma$). The overall four-sigma (very significant) discrepancy, therefore, is, in some sense, the product of a mediocre fit in the appearance channel, a poor fit in the disappearance channel, and the fact that the two subsets of data point to different regions of the parameter space.

The shapes observed in Fig.~\ref{Case2:T13T23Comp} are easy to understand. In a three-neutrino scheme, the disappearance probability $P_{\mu\mu}$ depends mostly on the $|U_{e3}|^2(1-|U_{\mu3}|^2)$. The best fit values translates into the relations $\sin^2\theta_{13}\sim 0.43(1+\sin^2\theta_{13})$ or $\sin^2\theta_{13}\sim 0.57(1+\sin^2\theta_{13})$, which explains the approximately linear shapes in Fig.~\ref{Case2:T13T23Comp}. The appearance channels, on the other hand, are mostly sensitive to the product $\sin^2{\theta_{13}}\sin^2{\theta_{23}}$, which explains the hyperbolic shape in Fig.~\ref{Case2:T13T23Comp}. Hence, in order to fit the four-neutrino data, the two different data sets wander towards different best-fit values for $\sin^2{\theta_{13}}$ and $\sin^2\theta_{23}$ as they strive to maintain the different combinations of these parameters constant. In Appendix~A, we repeat this analysis, this time simulating data consistent with the three-neutrino scenario. The results are depicted in Fig.~\ref{fig:3FlavorData_AppDis}. The shapes obtained from the two subsets are similar to those in Fig.~\ref{Case2:T13T23Comp}, but in this case the two analyses point to the same values of $\sin^2\theta_{13}$ and $\sin^2\theta_{23}$. 

In summary, not only is the goodness-of-fit poor, it is also possible to ascertain that different measurements of the mixing parameters are inconsistent with one another if one assumes that the three-massive-neutrinos paradigm is correct. There are several other ``inconsistency checks'' one would perform in order to reveal that new physics is affecting the long-baseline oscillations, including comparing data obtained with the neutrino beam and the antineutrino beam, comparing DUNE data with those from HyperKamiokande (same $L/E_{\nu}$ values, but different neutrino energies and baselines), comparing DUNE data with data from ``$\theta_{13}$'' reactor neutrino experiments \cite{Abe:2012tg,An:2012eh,Ahn:2012nd}, medium baseline reactor experiments \cite{Park:2014sja,Li:2014qca}, atmospheric neutrino experiments (for example, PINGU \cite{Aartsen:2014oha}), etc.

%Once the presence of new physics is established, a similar exercise will be required in order to figure out what is the new physics. Even if it turns out that a four-neutrino hypothesis provides a good fit to the data (as is, of course, the case here), it will take a significant amount of work (and data) to clearly establish what is going on. 

\section{Summary and Conclusions}
\label{conclusions}

Very ambitious next-generation long-baseline neutrino oscillation experiments are currently under serious consideration, especially the ``superbeam'' experiments Fermilab to DUNE in the United States and J-PARC to HyperKamiokande in Japan. Among the goals of these projects are searching for $CP$-invariance violation in the lepton sector and testing the limits of the three-massive-neutrinos paradigm. Here, we addressed the capabilities of the DUNE experiment to discover a fourth neutrino mass-eigenstate or, instead, constrain its existence, either falsifying or strengthening the three-massive-neutrinos paradigm. While several different new phenomena could manifest themselves at long-baseline neutrino experiments, we chose one new neutrino mass-eigenstate for a few reasons. First, oscillation effects due to a new light neutrino mass-eigenstate are easy to parameterize, and very familiar. Second, light sterile neutrinos are a very natural and benign extension of the Standard Model and could, for example, be a side effect of the mechanism responsible for the nonzero neutrino masses. Finally, the so-called short-baseline anomalies may be interpreted as evidence for new neutrino degrees of freedom so it is possible, even though the evidence is not very robust, that new neutrino states have already been found.
 
Assuming coherent oscillations, we discuss the oscillation probabilities involving a fourth neutrino for a wide range of values for the new mass-squared difference, including $|\Delta m^2_{14}|\gg |\Delta m^2_{13}|$, when the new oscillation length is very short and expected to lead to averaged-out effects at DUNE, $|\Delta m^2_{14}|\ll |\Delta m^2_{13}|$, when the new oscillation length is too long to be observed at DUNE, or  $|\Delta m^2_{14}|\sim|\Delta m^2_{13}|$, when DUNE is sensitive to the new oscillation frequency. We highlight the fact that, in all three cases, the values of the active elements of the fourth column of the mixing matrix, $U_{\alpha 4}$, $\alpha = e$, $\mu$, $\tau$, have a nontrivial impact on the experiment as long as these are large enough. We also discuss the extra sources of $CP$-invariance violation that arise from phases in the new mixing matrix elements. Given access to the $\nu_e$-appearance and $\nu_{\mu}$-disappearance channels, we find that DUNE is, in principle, sensitive to two of the three $CP$-odd phases in the mixing matrix.

We simulate data in the DUNE experiment assuming a 34 kt detector, a 1.2 MW proton beam, and 3 years each of neutrino and antineutrino data collection, exploring different scenarios. If the data are consistent with three-neutrinos (i.e., there are no accessible new light neutrinos) we find that DUNE is less sensitive than, for instance, the Daya Bay experiment when it comes to constraining $|U_{e4}|^2$ if $|\Delta m^2_{14}|\gtrsim|\Delta m^2_{13}|$, while DUNE can outperform current long-baseline experiments when it comes to constraining  $|U_{\mu4}|^2$ if $|\Delta m^2_{14}|\gtrsim|\Delta m^2_{13}|$. On the other hand, if $|\Delta m^2_{14}|\lesssim|\Delta m^2_{13}|$, DUNE outperforms all current experiments when it comes to constraining new, light neutrino mass-eigenstates thanks, in part, to the broad-band-beam nature of the experiment and the fact that it measures both $\nu_{\mu}$ disappearance and $\nu_e$ appearance. 

If the data are consistent with the existence of a fourth neutrino, DUNE has the capability to measure the new mixing parameters. This capability, however, depends strongly on the values of the parameters associated with the fourth neutrino, particularly $\Delta m_{14}^2$. We find that there are circumstances under which DUNE can not only discover new physics but also establish that there are new sources of $CP$-invariance violation. We emphasize that, if there is a new neutrino mass-eigenstate, the $\nu_e$ and $\nu_{\mu}$ data at DUNE can only explore a subset of the existing parameter space. One of the new mixing angles, and one of the two new sources of $CP$-invariance violation can only be accessed if one could also study $\nu_{\tau}$ appearance (or construct a $\nu_{\tau}$-beam, a much more challenging proposition).

Improved sensitivity is expected, of course, if backgrounds at DUNE turned out to be smaller than anticipated, if better energy resolution can be achieved, or if one were to optimize the beam energy profile. For example, we find that a $50\%$ background reduction yields $\mathcal{O}(10\%)$ improved sensitivity to a new neutrino state (e.g., $\mathcal{O}(10\%)$ stronger constraints in the $4|U_{e4}|^2|U_{\mu 4}|^2$ - $\Delta m_{14}^2$ plane, cf.~Fig.~\ref{SBLComp}). Access to higher neutrino energies, on the other hand, would allow DUNE more sensitivity to higher values of $\Delta m_{14}^2$.

We also briefly addressed whether DUNE data could reveal the existence of physics beyond the three-massive-neutrinos paradigm if the data were consistent with the existence of a fourth neutrino. We find that DUNE data are precise enough to reveal that a three-neutrino fit to data consistent with four neutrinos is poor (assuming the new mixing parameters are accessible). We also show that, in this scenario, fits to disjoint subsets of DUNE data point to different regions of the three-neutrino parameter space, another sign of new physics beyond three active neutrinos with nonzero mass. In order to properly diagnose that (a) there is physics beyond the three-massive-neutrinos paradigm, and (b) determine the nature of the new physics, it is very likely that one will need more/better data. We hope to return to these very important issues in a future study of long-baseline neutrino oscillations and new phenomena.

%%%%%%%%%%%%%
\section*{Acknowledgements}

This work is sponsored in part by the DOE grant \#DE-FG02-91ER40684. AK is also supported in part by the DOE grant  \#DE-SC0009919. 
%%%%%%%%%%%%%

\appendix
\setcounter{equation}{0}
\section{Three-neutrino Fits to Three-neutrino Data}

In this Appendix we present the results of simulating and analyzing DUNE data consistent with the three-neutrino hypothesis. This is done, in part, to validate our simulation and analysis tools, and in order to facilitate comparisons between the three-neutrino and the four-neutrino scenarios. We also comment on the ability of the DUNE experiment to constrain the solar parameters $\Delta m^2_{12}$ and $\theta_{12}$.

\begin{figure}[htbp]
	\begin{center}
	\includegraphics[width=0.95\linewidth]{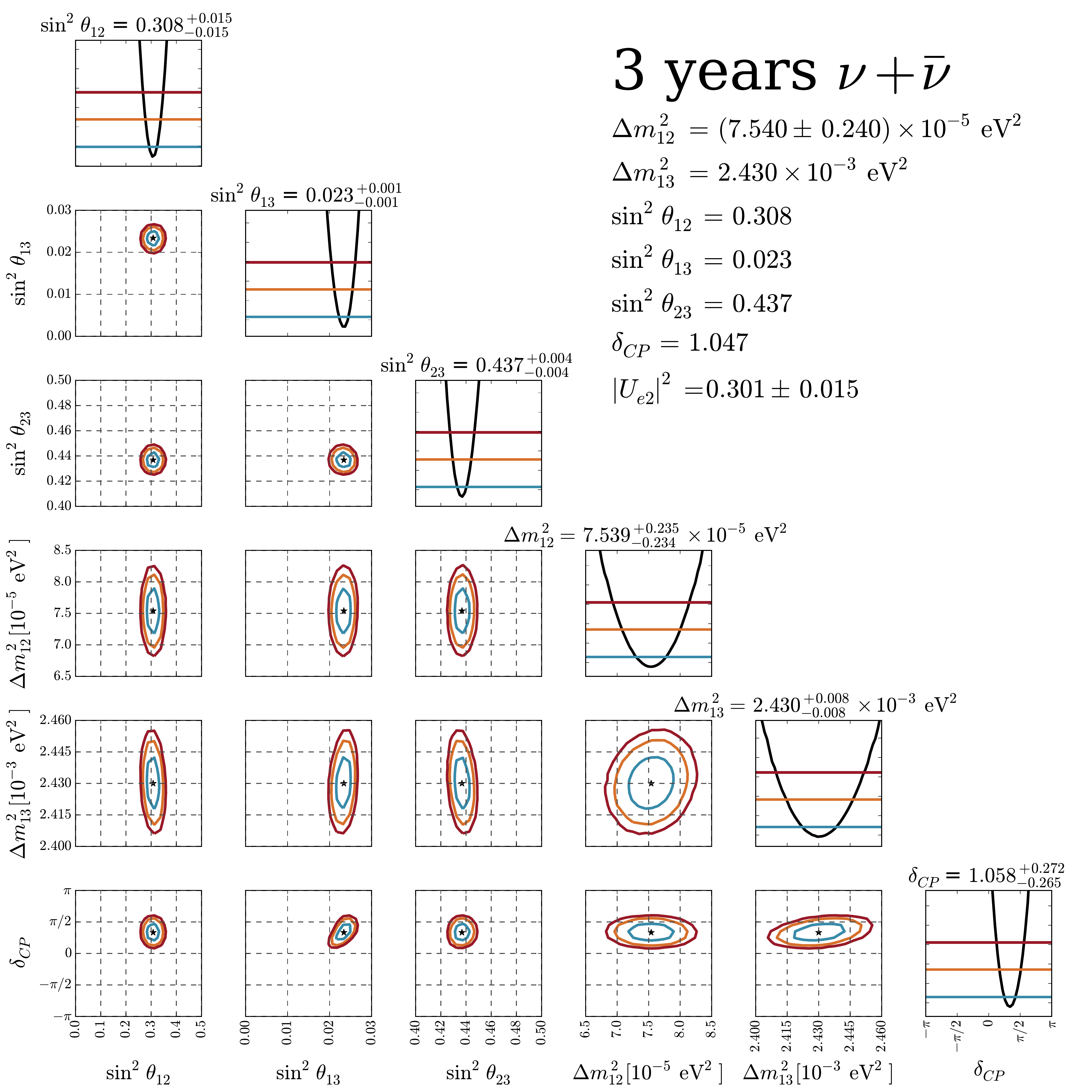}
	\caption{Expected sensitivity contours at 68.3\% (blue), 95\% (orange), and 99\% (red) CL as measured by DUNE with six years of data collection (3y $\nu ~+ $ 3y $\overline{\nu}$), a  34 kiloton detector, and a 1.2 MW beam given a three-neutrino scenario. On the far-right, one-dimensional $\Delta \chi^2$ plots for each parameter display the 68.3\% (blue), 95\% (orange), and 99\% (red) CL bounds. Quoted measurement bounds are for 68.3\% CL. Mixing angles here are $\sin^2\theta_{12} = 0.308$, $\sin^2\theta_{13} = 0.0235$, and $sin^2\theta_{23} = 0.437$. Results from solar neutrino experiments are included here as Gaussian priors for the values of $|U_{e2}|^2 = 0.301 \pm 0.015$ and $\Delta m_{12}^2 = (7.54\pm 0.24)\times 10^{-5}$ eV$^2$. This scenario assumes the normal hierarchy, i.e.\ $\Delta m_{13}^2 = +2.43\times 10^{-3}$ eV$^2$, and that $\delta_{CP} = \pi/3$.}
	\label{fig:3FlavorData_3FlavorFit}
	\end{center}
\end{figure}

Fig.~\ref{fig:3FlavorData_3FlavorFit} depicts the expected sensitivity contours at 68.3\% (blue), 95\% (orange), and 99\% (red) CL as measured by DUNE with six years of data collection (3 years with the neutrino beam, three years with the antineutrino one), a 34 kiloton detector, and a 1.2 MW beam, assuming the data are consistent with a three-neutrino scenario. The figure also depicts one-dimensional $\Delta \chi^2$ plots for each parameter, with the 68.3\% (blue), 95\% (orange), and 99\% (red) CL highlighted. Quoted measurement bounds are for 68.3\% CL. The input values of the mixing angles are $\sin^2\theta_{12} = 0.308$, $\sin^2\theta_{13} = 0.0235$, and $\sin^2\theta_{23} = 0.437$. Results from solar neutrino experiments and KamLAND are included here as Gaussian priors for the values of $|U_{e2}|^2 = 0.301 \pm 0.015$ and $\Delta m_{12}^2 = (7.54\pm 0.24)\times 10^{-5}$ eV$^2$. This scenario assumes a normal mass hierarchy, i.e.,\ $\Delta m_{13}^2 = +2.43\times 10^{-3}$ eV$^2$, and that the $CP$-odd phase is $\delta_{CP} = \pi/3$. Distributions are sampled using a Markov Chain Monte Carlo (MCMC) method~\cite{ForemanMackey:2012ig}. We find that the measurement precisions of the mixing angles and mass-squared differences are safely comparable to the projected results in Ref.~\cite{Adams:2013qkq}.

\begin{figure}[htbp]
	\begin{center}
	\includegraphics[width=0.7\linewidth]{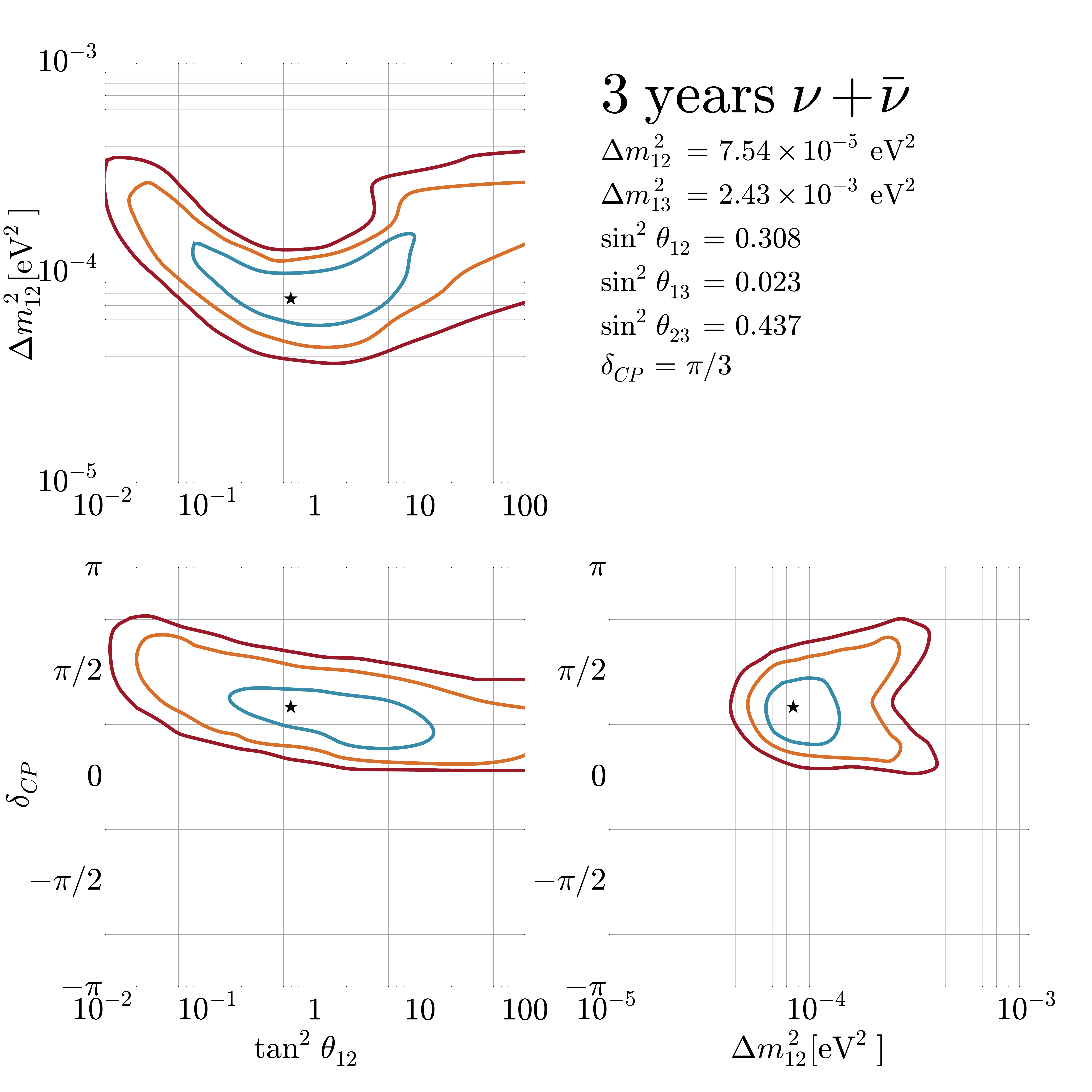}
	\caption{Expected sensitivity contours at 68.3\% (blue), 95\% (orange), and 99\% (red) as measured by DUNE with six years of data collection (3y $\nu ~+ $ 3y $\overline{\nu}$), a 34 kiloton detector, and a 1.2 MW beam assuming no fourth neutrino, and without external information from solar neutrino experiments included as Gaussian priors.}
	\label{fig:AppendixNoPrior}
	\end{center}
\end{figure}

Throughout, with one exception, our analyses include only data to be collected by the DUNE experiment. We do not include, for example, data from ``$\theta_{13}$'' reactor experiments,  \cite{Abe:2012tg,An:2012eh,Ahn:2012nd} nor do we include existing or simulated data from the long-baseline experiments currently in operation \cite{Ayres:2004js,Abe:2011ks,Abe:2012gx,Abe:2013hdq}. The main reason for this is that we anticipate DUNE data to provide the most significant information when it comes to measurements of $\sin^2\theta_{13}$, $\sin^2\theta_{23}$, $\Delta m^2_{13}$ (including the sign), and $\delta_{CP}$. We do, however, include results from solar data and from the KamLAND experiment when it comes to constraining the solar parameters $\Delta m^2_{12}$ and $\theta_{12}$. The reason for this is that DUNE's ability to, in isolation, determine the solar parameters is rather limited. To illustrate this fact, we perform a DUNE-only fit to the DUNE data. The results for a subset of the three-neutrino oscillation parameters ($\Delta m^2_{12}$, $\tan^2\theta_{12}$, $\delta_{CP}$) are depicted in Fig.~\ref{fig:AppendixNoPrior}.

While DUNE is not very sensitive to $\tan^2\theta_{12}$ or $\Delta m_{12}^2$ (note the logarithmic scales), it can exclude nonzero values for each parameter, and is still able to observe $CP$-invariance violation and measure $\delta_{CP}$ even if external information on the solar parameters is not included in the data analysis. The uncertainty on $\delta_{CP}$, as expected, is significantly larger (cf. Fig.~\ref{fig:3FlavorData_3FlavorFit}).

\begin{figure}[htbp]
	\begin{center}
	\includegraphics[width=0.45\linewidth]{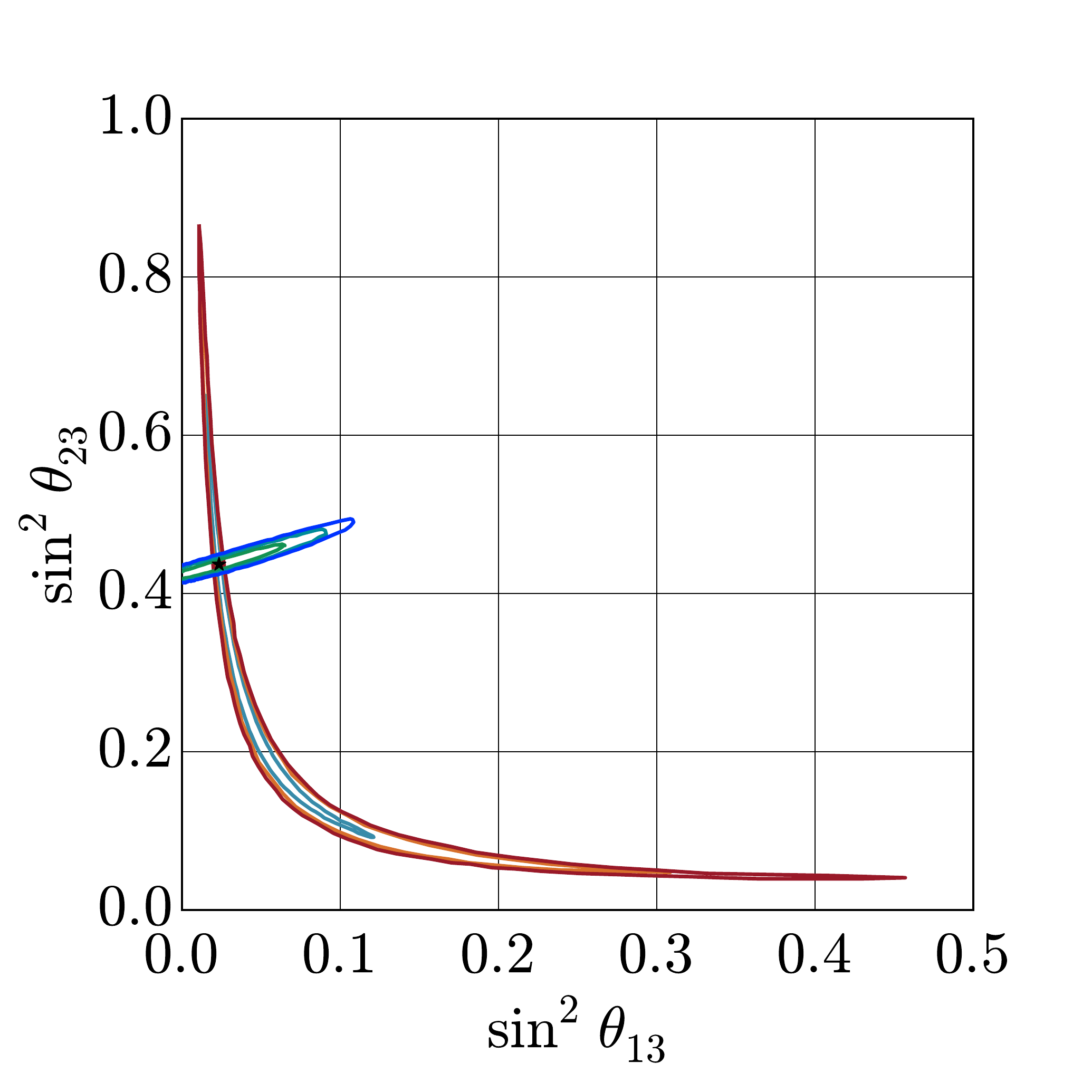}
	\caption{Expected sensitivity contours at 68.3\%, 95\%, and 99\% for neutrino and antineutrino appearance channels (blue, orange, red) vs.\ neutrino and antineutrino disappearance channels (green, teal, blue) in the $\sin^2{\theta_{13}}$ - $\sin^2{\theta_{23}}$ plane assuming a three neutrino hypothesis with parameters from Ref.~\cite{Agashe:2014kda}, indicated with a star in the figure.}
	\label{fig:3FlavorData_AppDis}
	\end{center}
\end{figure}

Finally, we repeat the analysis discussed in Sec.~\ref{subsec:Incorrect}, where appearance and disappearance channels are analyzed independently, this time assuming the data are consistent with the three-neutrino scenario. Fig.~\ref{fig:3FlavorData_AppDis} depicts the results. As in Sec.~\ref{subsec:Incorrect}, we see that the appearance channel is sensitive to the product $\sin^2{\theta_{13}}\sin^2{\theta_{23}}$ while the disappearance channel is mostly sensitive to $|U_{\mu3}|^2(1-|U_{\mu 3}|^2)$. Unlike the scenario discussed in Sec.~\ref{subsec:Incorrect} (see Fig.~\ref{Case2:T13T23Comp}), here the fits to the different data sets are in agreement.

\section{Four-neutrino Fits to Four-neutrino Data}

\begin{figure}[htbp]
	\begin{center}
	\includegraphics[width=1.0\linewidth]{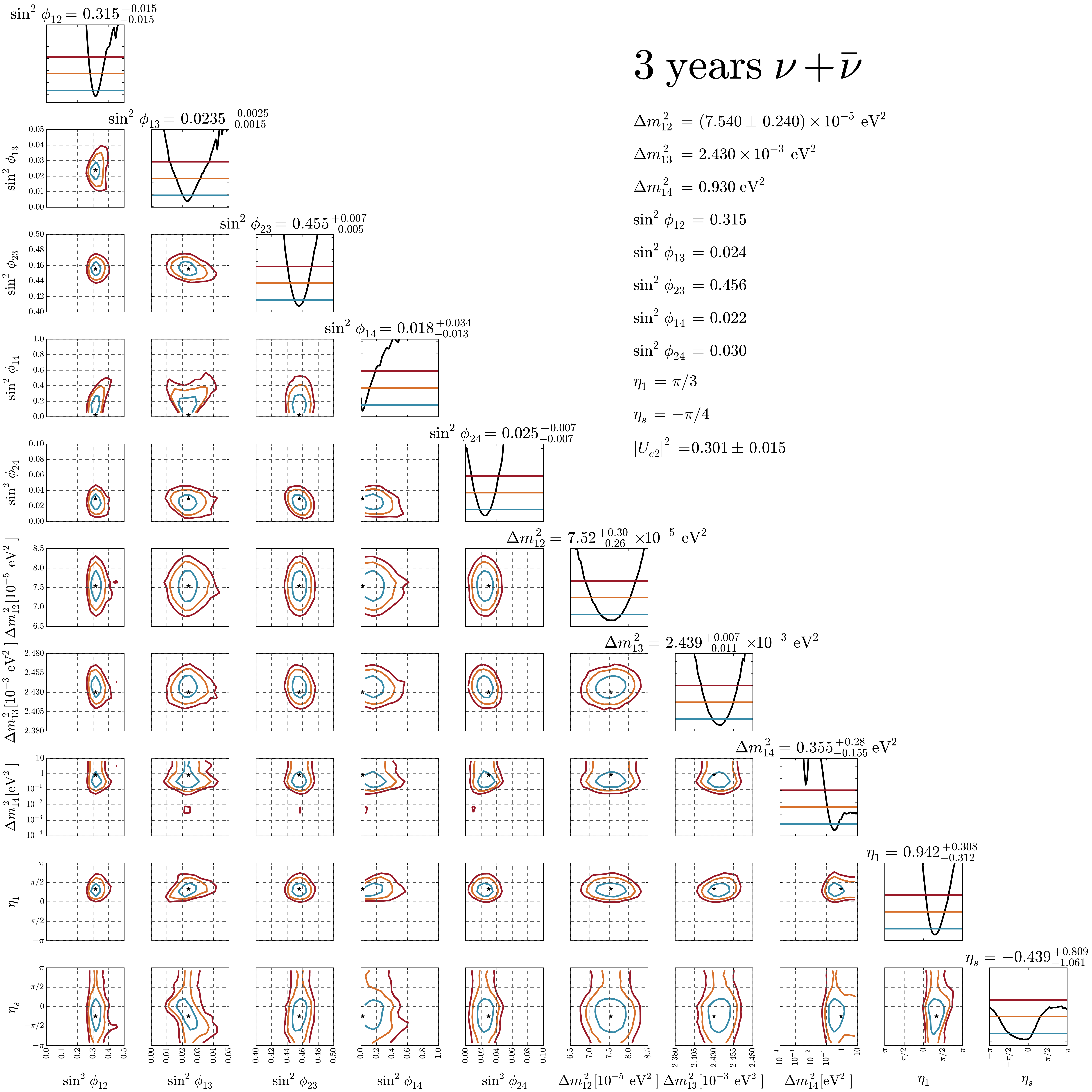}
	\caption{Expected sensitivity contours at 68.3\% (blue), 95\% (orange), and 99\% (red) CL at DUNE with six years of data collection (3y $\nu\ +$ 3y $\overline{\nu}$), a 34 kiloton detector, and a 1.2 MW beam given the existence of a fourth neutrino with parameters from Case 1 in Table~\ref{CaseTable}. On the far right, one-dimensional $\Delta \chi^2$ plots for each parameter display 68.3\% (blue), 95\% (orange), and 99\% (red) CL bounds. Quoted measurement bounds are for 68.3\% CL. Results from solar neutrino experiments are included here as Gaussian priors for the values of $|U_{e2}|^2 = 0.301 \pm 0.015$ and $\Delta m_{12}^2 = 7.54\pm 0.24\times 10^{-5}$ eV$^2$~\citep{Agashe:2014kda}.}
	\label{fig:AppendixCase1}
	\end{center}
\end{figure}

\begin{figure}[htbp]
	\begin{center}
	\includegraphics[width=1.0\linewidth]{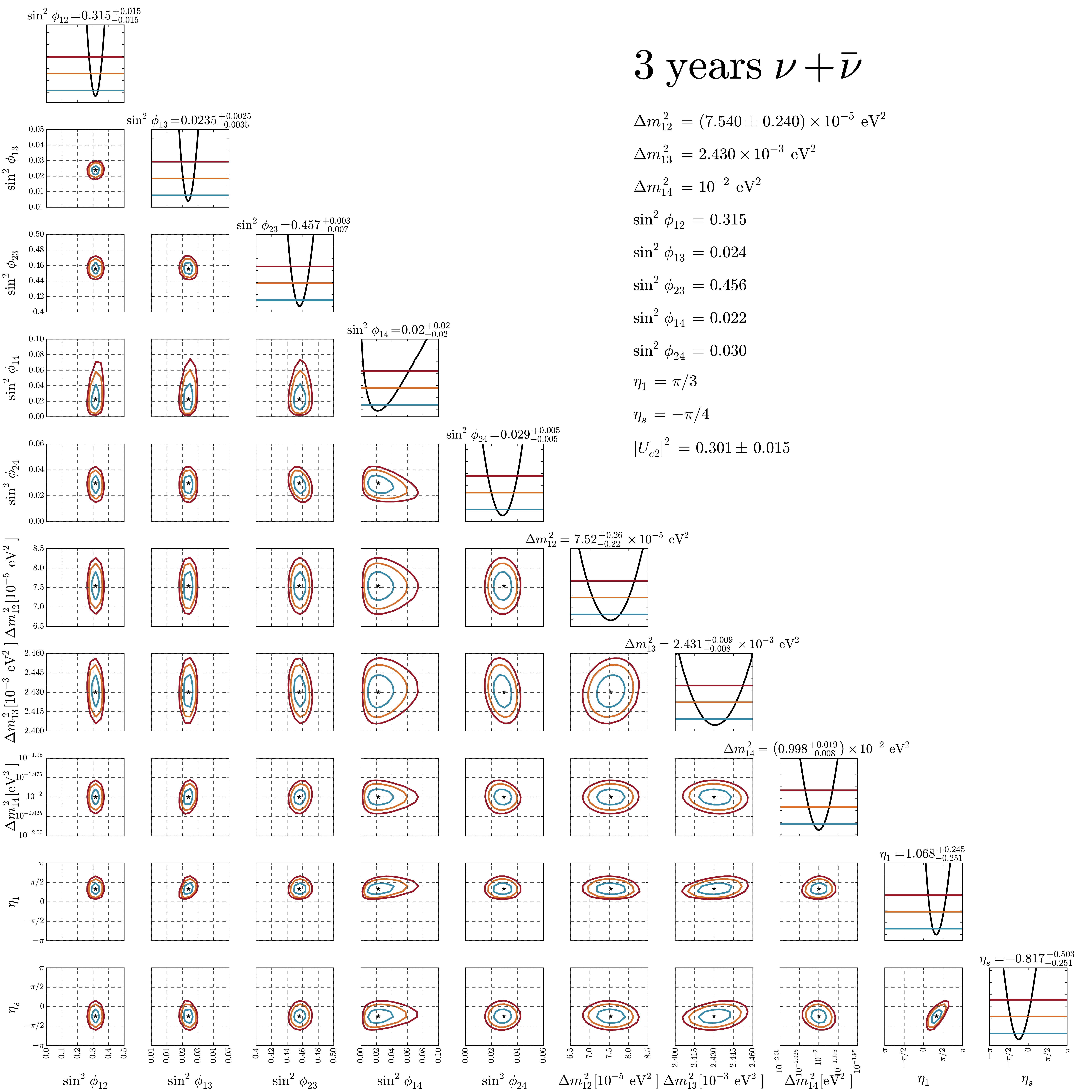}
	\caption{Expected sensitivity contours at 68.3\% (blue), 95\% (orange), and 99\% (red) CL at DUNE with six years of data collection (3y $\nu\ +$ 3y $\overline{\nu}$), a 34 kiloton detector, and a 1.2 MW beam given the existence of a fourth neutrino with parameters from Case 2 in Table~\ref{CaseTable}. On the far right, one-dimensional $\Delta \chi^2$ plots for each parameter display 68.3\% (blue), 95\% (orange), and 99\% (red) CL bounds. Quoted measurement bounds are for 68.3\% CL. Results from solar neutrino experiments are included here as Gaussian priors for the values of $|U_{e2}|^2 = 0.301 \pm 0.015$ and $\Delta m_{12}^2 = 7.54\pm 0.24\times 10^{-5}$ eV$^2$~\citep{Agashe:2014kda}.}
	\label{fig:DCPPi3_DSNPi4_KoppAngles_1en2Mass}
	\end{center}
\end{figure}

\begin{figure}[htbp]
	\begin{center}
	\includegraphics[width=1.0\linewidth]{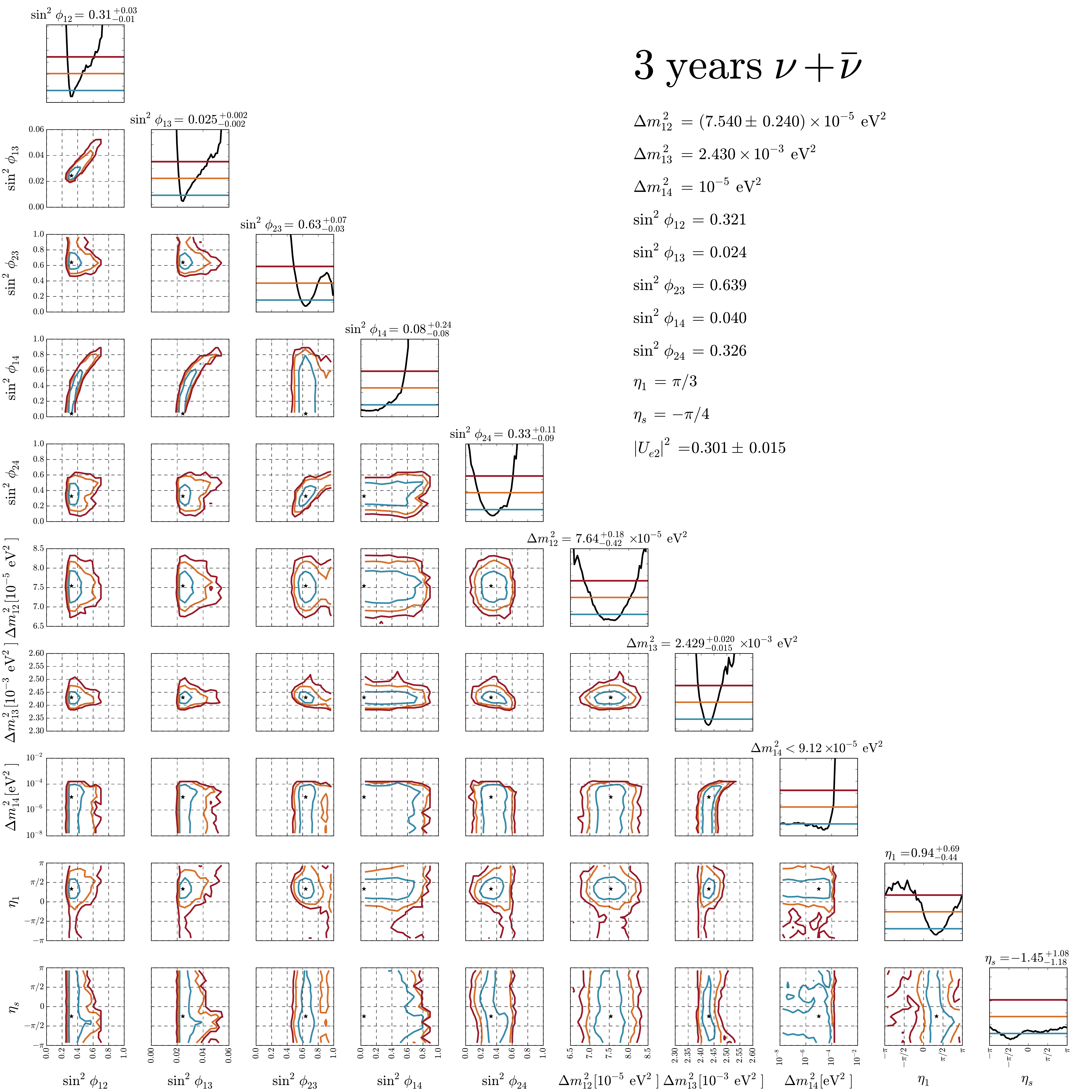}
	\caption{Expected sensitivity contours at 68.3\% (blue), 95\% (orange), and 99\% (red) CL at DUNE with six years of data collection (3y $\nu\ +$ 3y $\overline{\nu}$), a 34 kiloton detector, and a 1.2 MW beam given the existence of a fourth neutrino with parameters from Case 3 in Table~\ref{CaseTable}. On the far right, one-dimensional $\Delta \chi^2$ plots for each parameter display 68.3\% (blue), 95\% (orange), and 99\% (red) CL bounds.  Quoted measurement bounds are for 68.3\% CL. Results from solar neutrino experiments are included here as Gaussian priors for the values of $|U_{e2}|^2 = 0.301 \pm 0.015$ and $\Delta m_{12}^2 = 7.54\pm 0.24\times 10^{-5}$ eV$^2$~\citep{Agashe:2014kda}.}
	\label{fig:AppendixCase3}
	\end{center}
\end{figure}

Here we display the full results we obtain when analyzing the different four-neutrino scenarios (see Table~\ref{CaseTable}) assuming the four-neutrino hypothesis. Figs.~\ref{fig:AppendixCase1}, \ref{fig:DCPPi3_DSNPi4_KoppAngles_1en2Mass}, and \ref{fig:AppendixCase3} depict the expected sensitivity contours at 68.3\% (blue), 95\% (orange), and 99\% (red) CL at DUNE with six years of data collection (3y $\nu\ ~+$ 3y $\overline{\nu}$), a 34 kiloton detector, and a 1.2 MW beam, given the existence of a fourth neutrino with parameters from Case 1, Case 2, and Case 3 in Table~\ref{CaseTable}, respectively. Results from solar neutrino experiments are included here as Gaussian priors for the values of $|U_{e2}|^2 = 0.301 \pm 0.015$ and $\Delta m_{12}^2 = 7.54\pm 0.24\times 10^{-5}$ eV$^2$~\citep{Agashe:2014kda}. Distributions are sampled using a Markov Chain Monte Carlo (MCMC) method~\cite{ForemanMackey:2012ig}. These results are discussed in Sec.~\ref{subsec:MeasFourth}.

\bibliography{DUNEBib}{}

\begin{thebibliography}{45}
\expandafter\ifx\csname natexlab\endcsname\relax\def\natexlab#1{#1}\fi
\expandafter\ifx\csname bibnamefont\endcsname\relax
  \def\bibnamefont#1{#1}\fi
\expandafter\ifx\csname bibfnamefont\endcsname\relax
  \def\bibfnamefont#1{#1}\fi
\expandafter\ifx\csname citenamefont\endcsname\relax
  \def\citenamefont#1{#1}\fi
\expandafter\ifx\csname url\endcsname\relax
  \def\url#1{\texttt{#1}}\fi
\expandafter\ifx\csname urlprefix\endcsname\relax\def\urlprefix{URL }\fi
\providecommand{\bibinfo}[2]{#2}
\providecommand{\eprint}[2][]{\url{#2}}

\bibitem[{\citenamefont{Adams et~al.}(2013)}]{Adams:2013qkq}
\bibinfo{author}{\bibfnamefont{C.}~\bibnamefont{Adams}} \bibnamefont{et~al.}
  (\bibinfo{collaboration}{LBNE}) (\bibinfo{year}{2013}), \eprint{1307.7335}.

\bibitem[{\citenamefont{Kearns et~al.}(2013)}]{Kearns:2013lea}
\bibinfo{author}{\bibfnamefont{E.}~\bibnamefont{Kearns}} \bibnamefont{et~al.}
  (\bibinfo{collaboration}{Hyper-Kamiokande Working Group})
  (\bibinfo{year}{2013}), \eprint{1309.0184}.

\bibitem[{\citenamefont{de~Gouv\^ea et~al.}(2013)}]{deGouvea:2013onf}
\bibinfo{author}{\bibfnamefont{A.}~\bibnamefont{de~Gouv\^ea}}
  \bibnamefont{et~al.} (\bibinfo{collaboration}{Intensity Frontier Neutrino
  Working Group}) (\bibinfo{year}{2013}), \eprint{1310.4340}.

\bibitem[{\citenamefont{Qian et~al.}(2013)\citenamefont{Qian, Zhang, Diwan, and
  Vogel}}]{Qian:2013ora}
\bibinfo{author}{\bibfnamefont{X.}~\bibnamefont{Qian}},
  \bibinfo{author}{\bibfnamefont{C.}~\bibnamefont{Zhang}},
  \bibinfo{author}{\bibfnamefont{M.}~\bibnamefont{Diwan}}, \bibnamefont{and}
  \bibinfo{author}{\bibfnamefont{P.}~\bibnamefont{Vogel}}
  (\bibinfo{year}{2013}), \eprint{1308.5700}.

\bibitem[{\citenamefont{Ohlsson}(2013)}]{Ohlsson:2012kf}
\bibinfo{author}{\bibfnamefont{T.}~\bibnamefont{Ohlsson}},
  \bibinfo{journal}{Rept. Prog. Phys.} \textbf{\bibinfo{volume}{76}},
  \bibinfo{pages}{044201} (\bibinfo{year}{2013}), \eprint{1209.2710}.

\bibitem[{\citenamefont{Gonzalez-Garcia and
  Maltoni}(2008)}]{GonzalezGarcia:2007ib}
\bibinfo{author}{\bibfnamefont{M.}~\bibnamefont{Gonzalez-Garcia}}
  \bibnamefont{and} \bibinfo{author}{\bibfnamefont{M.}~\bibnamefont{Maltoni}},
  \bibinfo{journal}{Phys.~Rept.} \textbf{\bibinfo{volume}{460}},
  \bibinfo{pages}{1} (\bibinfo{year}{2008}), \eprint{0704.1800}.

\bibitem[{\citenamefont{Abazajian et~al.}(2012)\citenamefont{Abazajian, Acero,
  Agarwalla, Aguilar-Arevalo, Albright et~al.}}]{Abazajian:2012ys}
\bibinfo{author}{\bibfnamefont{K.}~\bibnamefont{Abazajian}},
  \bibinfo{author}{\bibfnamefont{M.}~\bibnamefont{Acero}},
  \bibinfo{author}{\bibfnamefont{S.}~\bibnamefont{Agarwalla}},
  \bibinfo{author}{\bibfnamefont{A.}~\bibnamefont{Aguilar-Arevalo}},
  \bibinfo{author}{\bibfnamefont{C.}~\bibnamefont{Albright}},
  \bibnamefont{et~al.} (\bibinfo{year}{2012}), \eprint{1204.5379}.

\bibitem[{\citenamefont{Donini et~al.}(2007)\citenamefont{Donini, Maltoni,
  Meloni, Migliozzi, and Terranova}}]{Donini:2007yf}
\bibinfo{author}{\bibfnamefont{A.}~\bibnamefont{Donini}},
  \bibinfo{author}{\bibfnamefont{M.}~\bibnamefont{Maltoni}},
  \bibinfo{author}{\bibfnamefont{D.}~\bibnamefont{Meloni}},
  \bibinfo{author}{\bibfnamefont{P.}~\bibnamefont{Migliozzi}},
  \bibnamefont{and}
  \bibinfo{author}{\bibfnamefont{F.}~\bibnamefont{Terranova}},
  \bibinfo{journal}{JHEP} \textbf{\bibinfo{volume}{12}}, \bibinfo{pages}{013}
  (\bibinfo{year}{2007}), \eprint{0704.0388}.

\bibitem[{\citenamefont{Dighe and Ray}(2007)}]{Dighe:2007uf}
\bibinfo{author}{\bibfnamefont{A.}~\bibnamefont{Dighe}} \bibnamefont{and}
  \bibinfo{author}{\bibfnamefont{S.}~\bibnamefont{Ray}},
  \bibinfo{journal}{Phys. Rev.} \textbf{\bibinfo{volume}{D76}},
  \bibinfo{pages}{113001} (\bibinfo{year}{2007}), \eprint{0709.0383}.

\bibitem[{\citenamefont{de~Gouv\^ea and Wytock}(2009)}]{deGouvea:2008qk}
\bibinfo{author}{\bibfnamefont{A.}~\bibnamefont{de~Gouv\^ea}} \bibnamefont{and}
  \bibinfo{author}{\bibfnamefont{T.}~\bibnamefont{Wytock}},
  \bibinfo{journal}{Phys. Rev.} \textbf{\bibinfo{volume}{D79}},
  \bibinfo{pages}{073005} (\bibinfo{year}{2009}), \eprint{0809.5076}.

\bibitem[{\citenamefont{Meloni et~al.}(2010)\citenamefont{Meloni, Tang, and
  Winter}}]{Meloni:2010zr}
\bibinfo{author}{\bibfnamefont{D.}~\bibnamefont{Meloni}},
  \bibinfo{author}{\bibfnamefont{J.}~\bibnamefont{Tang}}, \bibnamefont{and}
  \bibinfo{author}{\bibfnamefont{W.}~\bibnamefont{Winter}},
  \bibinfo{journal}{Phys. Rev.} \textbf{\bibinfo{volume}{D82}},
  \bibinfo{pages}{093008} (\bibinfo{year}{2010}), \eprint{1007.2419}.

\bibitem[{\citenamefont{Bhattacharya et~al.}(2012)\citenamefont{Bhattacharya,
  Thalapillil, and Wagner}}]{Bhattacharya:2011ee}
\bibinfo{author}{\bibfnamefont{B.}~\bibnamefont{Bhattacharya}},
  \bibinfo{author}{\bibfnamefont{A.~M.} \bibnamefont{Thalapillil}},
  \bibnamefont{and} \bibinfo{author}{\bibfnamefont{C.~E.~M.}
  \bibnamefont{Wagner}}, \bibinfo{journal}{Phys. Rev.}
  \textbf{\bibinfo{volume}{D85}}, \bibinfo{pages}{073004}
  (\bibinfo{year}{2012}), \eprint{1111.4225}.

\bibitem[{\citenamefont{Hollander and Mocioiu}(2014)}]{Hollander:2014iha}
\bibinfo{author}{\bibfnamefont{D.}~\bibnamefont{Hollander}} \bibnamefont{and}
  \bibinfo{author}{\bibfnamefont{I.}~\bibnamefont{Mocioiu}}
  (\bibinfo{year}{2014}), \eprint{1408.1749}.

\bibitem[{\citenamefont{Klop and Palazzo}(2014)}]{Klop:2014ima}
\bibinfo{author}{\bibfnamefont{N.}~\bibnamefont{Klop}} \bibnamefont{and}
  \bibinfo{author}{\bibfnamefont{A.}~\bibnamefont{Palazzo}}
  (\bibinfo{year}{2014}), \eprint{1412.7524}.

\bibitem[{\citenamefont{de~Gouv\^ea}(2005)}]{deGouvea:2005er}
\bibinfo{author}{\bibfnamefont{A.}~\bibnamefont{de~Gouv\^ea}},
  \bibinfo{journal}{Phys.~Rev.} \textbf{\bibinfo{volume}{D72}},
  \bibinfo{pages}{033005} (\bibinfo{year}{2005}), \eprint{hep-ph/0501039}.

\bibitem[{\citenamefont{Asaka et~al.}(2005)\citenamefont{Asaka, Blanchet, and
  Shaposhnikov}}]{Asaka:2005an}
\bibinfo{author}{\bibfnamefont{T.}~\bibnamefont{Asaka}},
  \bibinfo{author}{\bibfnamefont{S.}~\bibnamefont{Blanchet}}, \bibnamefont{and}
  \bibinfo{author}{\bibfnamefont{M.}~\bibnamefont{Shaposhnikov}},
  \bibinfo{journal}{Phys.~Lett.} \textbf{\bibinfo{volume}{B631}},
  \bibinfo{pages}{151} (\bibinfo{year}{2005}), \eprint{hep-ph/0503065}.

\bibitem[{\citenamefont{Aguilar-Arevalo et~al.}(2001)}]{Aguilar:2001ty}
\bibinfo{author}{\bibfnamefont{A.}~\bibnamefont{Aguilar-Arevalo}}
  \bibnamefont{et~al.} (\bibinfo{collaboration}{LSND Collaboration}),
  \bibinfo{journal}{Phys.~Rev.} \textbf{\bibinfo{volume}{D64}},
  \bibinfo{pages}{112007} (\bibinfo{year}{2001}), \eprint{hep-ex/0104049}.

\bibitem[{\citenamefont{Aguilar-Arevalo et~al.}(2009)}]{AguilarArevalo:2008rc}
\bibinfo{author}{\bibfnamefont{A.}~\bibnamefont{Aguilar-Arevalo}}
  \bibnamefont{et~al.} (\bibinfo{collaboration}{MiniBooNE Collaboration}),
  \bibinfo{journal}{Phys.~Rev.~Lett.} \textbf{\bibinfo{volume}{102}},
  \bibinfo{pages}{101802} (\bibinfo{year}{2009}), \eprint{0812.2243}.

\bibitem[{\citenamefont{Mention et~al.}(2011)\citenamefont{Mention, Fechner,
  Lasserre, Mueller, Lhuillier et~al.}}]{Mention:2011rk}
\bibinfo{author}{\bibfnamefont{G.}~\bibnamefont{Mention}},
  \bibinfo{author}{\bibfnamefont{M.}~\bibnamefont{Fechner}},
  \bibinfo{author}{\bibfnamefont{T.}~\bibnamefont{Lasserre}},
  \bibinfo{author}{\bibfnamefont{T.}~\bibnamefont{Mueller}},
  \bibinfo{author}{\bibfnamefont{D.}~\bibnamefont{Lhuillier}},
  \bibnamefont{et~al.}, \bibinfo{journal}{Phys.~Rev.}
  \textbf{\bibinfo{volume}{D83}}, \bibinfo{pages}{073006}
  (\bibinfo{year}{2011}), \eprint{1101.2755}.

\bibitem[{\citenamefont{Frekers et~al.}(2011)\citenamefont{Frekers, Ejiri,
  Akimune, Adachi, Bilgier et~al.}}]{Frekers:2011zz}
\bibinfo{author}{\bibfnamefont{D.}~\bibnamefont{Frekers}},
  \bibinfo{author}{\bibfnamefont{H.}~\bibnamefont{Ejiri}},
  \bibinfo{author}{\bibfnamefont{H.}~\bibnamefont{Akimune}},
  \bibinfo{author}{\bibfnamefont{T.}~\bibnamefont{Adachi}},
  \bibinfo{author}{\bibfnamefont{B.}~\bibnamefont{Bilgier}},
  \bibnamefont{et~al.}, \bibinfo{journal}{Phys.~Lett.}
  \textbf{\bibinfo{volume}{B706}}, \bibinfo{pages}{134} (\bibinfo{year}{2011}).

\bibitem[{\citenamefont{Aguilar-Arevalo
  et~al.}(2013)}]{Aguilar-Arevalo:2013pmq}
\bibinfo{author}{\bibfnamefont{A.}~\bibnamefont{Aguilar-Arevalo}}
  \bibnamefont{et~al.} (\bibinfo{collaboration}{MiniBooNE Collaboration}),
  \bibinfo{journal}{Phys.~Rev.~Lett.} \textbf{\bibinfo{volume}{110}},
  \bibinfo{pages}{161801} (\bibinfo{year}{2013}), \eprint{1207.4809}.

\bibitem[{\citenamefont{Olive et~al.}(2014)}]{Agashe:2014kda}
\bibinfo{author}{\bibfnamefont{K.}~\bibnamefont{Olive}} \bibnamefont{et~al.}
  (\bibinfo{collaboration}{Particle Data Group}),
  \bibinfo{journal}{Chin.~Phys.} \textbf{\bibinfo{volume}{C38}},
  \bibinfo{pages}{090001} (\bibinfo{year}{2014}).

\bibitem[{\citenamefont{Albright et~al.}(2000)\citenamefont{Albright, Anderson,
  Barger, Bernstein, Blazey et~al.}}]{Albright:2000xi}
\bibinfo{author}{\bibfnamefont{C.}~\bibnamefont{Albright}},
  \bibinfo{author}{\bibfnamefont{G.}~\bibnamefont{Anderson}},
  \bibinfo{author}{\bibfnamefont{V.}~\bibnamefont{Barger}},
  \bibinfo{author}{\bibfnamefont{R.}~\bibnamefont{Bernstein}},
  \bibinfo{author}{\bibfnamefont{G.}~\bibnamefont{Blazey}},
  \bibnamefont{et~al.} (\bibinfo{year}{2000}), \eprint{hep-ex/0008064}.

\bibitem[{\citenamefont{Formaggio and Zeller}(2012)}]{Formaggio:2013kya}
\bibinfo{author}{\bibfnamefont{J.}~\bibnamefont{Formaggio}} \bibnamefont{and}
  \bibinfo{author}{\bibfnamefont{G.}~\bibnamefont{Zeller}},
  \bibinfo{journal}{Rev.~Mod.~Phys.} \textbf{\bibinfo{volume}{84}},
  \bibinfo{pages}{1307} (\bibinfo{year}{2012}), \eprint{1305.7513}.

\bibitem[{\citenamefont{An et~al.}(2014)}]{An:2014bik}
\bibinfo{author}{\bibfnamefont{F.}~\bibnamefont{An}} \bibnamefont{et~al.}
  (\bibinfo{collaboration}{Daya Bay}), \bibinfo{journal}{Phys.Rev.Lett.}
  \textbf{\bibinfo{volume}{113}}, \bibinfo{pages}{141802}
  (\bibinfo{year}{2014}), \eprint{1407.7259}.

\bibitem[{\citenamefont{Declais et~al.}(1995)\citenamefont{Declais, Favier,
  Metref, Pessard, Achkar et~al.}}]{Declais:1994su}
\bibinfo{author}{\bibfnamefont{Y.}~\bibnamefont{Declais}},
  \bibinfo{author}{\bibfnamefont{J.}~\bibnamefont{Favier}},
  \bibinfo{author}{\bibfnamefont{A.}~\bibnamefont{Metref}},
  \bibinfo{author}{\bibfnamefont{H.}~\bibnamefont{Pessard}},
  \bibinfo{author}{\bibfnamefont{B.}~\bibnamefont{Achkar}},
  \bibnamefont{et~al.}, \bibinfo{journal}{Nucl.Phys.}
  \textbf{\bibinfo{volume}{B434}}, \bibinfo{pages}{503} (\bibinfo{year}{1995}).

\bibitem[{\citenamefont{Timmons}(2015)}]{Timmons:2015lga}
\bibinfo{author}{\bibfnamefont{A.}~\bibnamefont{Timmons}}
  (\bibinfo{year}{2015}), \eprint{1504.04046}.

\bibitem[{\citenamefont{Adey et~al.}(2014)}]{Adey:2014rfv}
\bibinfo{author}{\bibfnamefont{D.}~\bibnamefont{Adey}} \bibnamefont{et~al.}
  (\bibinfo{collaboration}{nuSTORM Collaboration}),
  \bibinfo{journal}{Phys.~Rev.} \textbf{\bibinfo{volume}{D89}},
  \bibinfo{pages}{071301} (\bibinfo{year}{2014}), \eprint{1402.5250}.

\bibitem[{\citenamefont{de~Gouv\^ea et~al.}(2015)\citenamefont{de~Gouv\^ea,
  Kelly, and Kobach}}]{deGouvea:2014aoa}
\bibinfo{author}{\bibfnamefont{A.}~\bibnamefont{de~Gouv\^ea}},
  \bibinfo{author}{\bibfnamefont{K.~J.} \bibnamefont{Kelly}}, \bibnamefont{and}
  \bibinfo{author}{\bibfnamefont{A.}~\bibnamefont{Kobach}},
  \bibinfo{journal}{Phys.~Rev.} \textbf{\bibinfo{volume}{D91}},
  \bibinfo{pages}{053005} (\bibinfo{year}{2015}), \eprint{1412.1479}.

\bibitem[{\citenamefont{Agafonova et~al.}(2013)}]{Agafonova:2013xsk}
\bibinfo{author}{\bibfnamefont{N.}~\bibnamefont{Agafonova}}
  \bibnamefont{et~al.} (\bibinfo{collaboration}{OPERA}),
  \bibinfo{journal}{JHEP} \textbf{\bibinfo{volume}{1307}}, \bibinfo{pages}{004}
  (\bibinfo{year}{2013}), \eprint{1303.3953}.

\bibitem[{\citenamefont{Antonello et~al.}(2013)}]{Antonello:2013gut}
\bibinfo{author}{\bibfnamefont{M.}~\bibnamefont{Antonello}}
  \bibnamefont{et~al.} (\bibinfo{collaboration}{ICARUS}),
  \bibinfo{journal}{Eur.~Phys.~J.} \textbf{\bibinfo{volume}{C73}},
  \bibinfo{pages}{2599} (\bibinfo{year}{2013}), \eprint{1307.4699}.

\bibitem[{\citenamefont{Kyberd et~al.}(2012)}]{Kyberd:2012iz}
\bibinfo{author}{\bibfnamefont{P.}~\bibnamefont{Kyberd}} \bibnamefont{et~al.}
  (\bibinfo{collaboration}{nuSTORM Collaboration}) (\bibinfo{year}{2012}),
  \eprint{1206.0294}.

\bibitem[{\citenamefont{Kopp et~al.}(2013)\citenamefont{Kopp, Machado, Maltoni,
  and Schwetz}}]{Kopp:2013vaa}
\bibinfo{author}{\bibfnamefont{J.}~\bibnamefont{Kopp}},
  \bibinfo{author}{\bibfnamefont{P.~A.~N.} \bibnamefont{Machado}},
  \bibinfo{author}{\bibfnamefont{M.}~\bibnamefont{Maltoni}}, \bibnamefont{and}
  \bibinfo{author}{\bibfnamefont{T.}~\bibnamefont{Schwetz}},
  \bibinfo{journal}{JHEP} \textbf{\bibinfo{volume}{1305}}, \bibinfo{pages}{050}
  (\bibinfo{year}{2013}), \eprint{1303.3011}.

\bibitem[{\citenamefont{Palazzo}(2012)}]{Palazzo:2012yf}
\bibinfo{author}{\bibfnamefont{A.}~\bibnamefont{Palazzo}},
  \bibinfo{journal}{Phys. Rev.} \textbf{\bibinfo{volume}{D85}},
  \bibinfo{pages}{077301} (\bibinfo{year}{2012}), \eprint{1201.4280}.

\bibitem[{\citenamefont{Foreman-Mackey
  et~al.}(2013)\citenamefont{Foreman-Mackey, Hogg, Lang, and
  Goodman}}]{ForemanMackey:2012ig}
\bibinfo{author}{\bibfnamefont{D.}~\bibnamefont{Foreman-Mackey}},
  \bibinfo{author}{\bibfnamefont{D.~W.} \bibnamefont{Hogg}},
  \bibinfo{author}{\bibfnamefont{D.}~\bibnamefont{Lang}}, \bibnamefont{and}
  \bibinfo{author}{\bibfnamefont{J.}~\bibnamefont{Goodman}},
  \bibinfo{journal}{Publ.~Astron.~Soc.~Pacific} \textbf{\bibinfo{volume}{125}},
  \bibinfo{pages}{306} (\bibinfo{year}{2013}), \eprint{1202.3665}.

\bibitem[{\citenamefont{Abe et~al.}(2012{\natexlab{a}})}]{Abe:2012tg}
\bibinfo{author}{\bibfnamefont{Y.}~\bibnamefont{Abe}} \bibnamefont{et~al.}
  (\bibinfo{collaboration}{Double Chooz}), \bibinfo{journal}{Phys.~Rev.}
  \textbf{\bibinfo{volume}{D86}}, \bibinfo{pages}{052008}
  (\bibinfo{year}{2012}{\natexlab{a}}), \eprint{1207.6632}.

\bibitem[{\citenamefont{An et~al.}(2012)}]{An:2012eh}
\bibinfo{author}{\bibfnamefont{F.}~\bibnamefont{An}} \bibnamefont{et~al.}
  (\bibinfo{collaboration}{Daya Bay}), \bibinfo{journal}{Phys.~Rev.~Lett.}
  \textbf{\bibinfo{volume}{108}}, \bibinfo{pages}{171803}
  (\bibinfo{year}{2012}), \eprint{1203.1669}.

\bibitem[{\citenamefont{Ahn et~al.}(2012)}]{Ahn:2012nd}
\bibinfo{author}{\bibfnamefont{J.}~\bibnamefont{Ahn}} \bibnamefont{et~al.}
  (\bibinfo{collaboration}{RENO}), \bibinfo{journal}{Phys.~Rev.~Lett.}
  \textbf{\bibinfo{volume}{108}}, \bibinfo{pages}{191802}
  (\bibinfo{year}{2012}), \eprint{1204.0626}.

\bibitem[{\citenamefont{Park}(2013)}]{Park:2014sja}
\bibinfo{author}{\bibfnamefont{J.}~\bibnamefont{Park}}, \bibinfo{journal}{PoS}
  \textbf{\bibinfo{volume}{Neutel2013}}, \bibinfo{pages}{076}
  (\bibinfo{year}{2013}).

\bibitem[{\citenamefont{Li}(2014)}]{Li:2014qca}
\bibinfo{author}{\bibfnamefont{Y.-F.} \bibnamefont{Li}},
  \bibinfo{journal}{Int.~J.~Mod.~Phys.~Conf.~Ser.}
  \textbf{\bibinfo{volume}{31}}, \bibinfo{pages}{1460300}
  (\bibinfo{year}{2014}), \eprint{1402.6143}.

\bibitem[{\citenamefont{Aartsen et~al.}(2014)}]{Aartsen:2014oha}
\bibinfo{author}{\bibfnamefont{M.}~\bibnamefont{Aartsen}} \bibnamefont{et~al.}
  (\bibinfo{collaboration}{IceCube PINGU}) (\bibinfo{year}{2014}),
  \eprint{1401.2046}.

\bibitem[{\citenamefont{Ayres et~al.}(2004)}]{Ayres:2004js}
\bibinfo{author}{\bibfnamefont{D.}~\bibnamefont{Ayres}} \bibnamefont{et~al.}
  (\bibinfo{collaboration}{NOvA}) (\bibinfo{year}{2004}),
  \eprint{hep-ex/0503053}.

\bibitem[{\citenamefont{Abe et~al.}(2011)}]{Abe:2011ks}
\bibinfo{author}{\bibfnamefont{K.}~\bibnamefont{Abe}} \bibnamefont{et~al.}
  (\bibinfo{collaboration}{T2K}), \bibinfo{journal}{Nucl. Instrum. Meth.}
  \textbf{\bibinfo{volume}{A659}}, \bibinfo{pages}{106} (\bibinfo{year}{2011}),
  \eprint{1106.1238}.

\bibitem[{\citenamefont{Abe et~al.}(2012{\natexlab{b}})}]{Abe:2012gx}
\bibinfo{author}{\bibfnamefont{K.}~\bibnamefont{Abe}} \bibnamefont{et~al.}
  (\bibinfo{collaboration}{T2K}), \bibinfo{journal}{Phys. Rev.}
  \textbf{\bibinfo{volume}{D85}}, \bibinfo{pages}{031103}
  (\bibinfo{year}{2012}{\natexlab{b}}), \eprint{1201.1386}.

\bibitem[{\citenamefont{Abe et~al.}(2014)}]{Abe:2013hdq}
\bibinfo{author}{\bibfnamefont{K.}~\bibnamefont{Abe}} \bibnamefont{et~al.}
  (\bibinfo{collaboration}{T2K}), \bibinfo{journal}{Phys. Rev. Lett.}
  \textbf{\bibinfo{volume}{112}}, \bibinfo{pages}{061802}
  (\bibinfo{year}{2014}), \eprint{1311.4750}.

\end{thebibliography}

\end{document}